\documentclass[twocolumn,showpacs,preprintnumbers,amsmath,amssymb,floatfix]{revtex4}

\usepackage{amsmath}
\usepackage{natbib}
\usepackage{graphicx}
\usepackage{subfigure}
\usepackage{mathrsfs}
\usepackage{booktabs}
\usepackage{siunitx}
\usepackage{dcolumn}
\usepackage{bm}
\usepackage{color}
\usepackage[section]{placeins}
\usepackage{hyperref}
\begin{document}
\title{Light Curves of Chaotic Charged Hot-Spots in Curved Spacetime: Opening an Observational Window to Chaos}
\author{Shiyang Hu$^{1}$}
\email{husy\_arcturus@163.com}
\author{Dan Li$^{1}$}
\author{Chen Deng$^{2}$}
\affiliation{1. School of Mathematics and Physics, University of South China, Hengyang 421001, People's Republic of China \\
             2. School of Astronomy and Space Science, Nanjing University, Nanjing 210023, People's Republic of China}
\begin{abstract}
The observed scarcity of chaotic phenomena in astronomy contrasts sharply with their theoretical significance, primarily due to the absence of a robust framework for detecting chaos. In this study, we numerically simulate the light curves of hot-spots in Kerr spacetime under the influence of an external asymptotically uniform electromagnetic field. Our results reveal a clear distinction between the light curves of chaotic and regular hot-spots, particularly in their power spectra: the latter display isolated, sharp peaks, while the former exhibit broad, continuous peaks of low amplitude. These findings highlight the potential of using light curves as a probe for chaotic orbits in curved spacetime.
\end{abstract}
\keywords{Curved spacetime; Light curves; Chaotic orbits; Hot-spot; Ray-tracing method}
\maketitle
\emph{Introduction.}---A defining characteristic of chaotic behavior is that infinitesimal perturbations in the initial conditions of a dynamical system can lead to vastly divergent evolutionary outcomes. Chaos is pervasive in astronomy \cite{Levin (2000),King and Pringle (2006),Takahashi and Koyama (2009),Kopacek et al. (2010),Kopacek et al. (2014),Cunha et al. (2016)}, and substantial numerical evidence suggests that chaotic orbits may play a crucial role in explaining high-energy astrophysical phenomena, such as the polarization degree of fast radio bursts \cite{Wang et al. (2021)}, particle acceleration and the formation of relativistic jets \cite{Stuchlik and Kolos (2016)}. However, no comprehensive method currently exists for detecting chaos in celestial dynamical systems.

As early as $1988$, Sussman and Wisdom employed a 12th-order St\"{o}rmer predictor to numerically simulate the motion of trans-Neptunian objects, revealing that Pluto's maximum Lyapunov exponent diverges on a timescale of approximately 20 million years, indicating its chaotic nature \cite{Sussman and Wisdom (1988)}. This finding was later corroborated by Wisdom and Holman \cite{Wisdom and Holman (1991)}. However, such chaotic dynamics remain beyond our observational reach, as they are inherently long-term phenomenon.

In dynamical simulations of post-Newtonian spinning compact binaries, several authors have identified chaotic orbits and explored their correlations with binary spins, mass ratio, mass quadrupole moment, and orbital eccentricity \cite{Levin (2003),Cornish and Levin (2003),Hartl and Buonanno (2005),Wu and Xie (2008),Mei et al. (2013),Li et al. (2019),Hu et al. (2021)}. However, since the chaotic timescale of binaries exceeds the inspiral timescale, laser interferometer gravitational-wave observatories (LIGO), designed to detect binary coalescences, are unable to capture signatures of chaos \cite{Schnittman and Rasio (2001)}. Although Destounis et al. suggested that the gravitational-wave frequency in chaotic extreme mass-ratio inspiral (EMRI) systems may exhibit abrupt increases \cite{Destounis et al. (2021)}, such observations will have to await the successful deployment of space-based gravitational-wave interferometers. Currently, identifying reliable methods to detect chaotic orbits remains an open question.

In black hole spacetime, a timelike particle orbiting the black hole continuously emits electromagnetic radiation, allowing the particle to be treated as a ``hot-spot''. The dynamical system comprising a black hole and hot-spot has been shown to effectively explain flares and light curves observed near the supermassive black hole at the center of the Milky Way \cite{Genzel et al. (2003),Baubock et al. (2020),Matsumoto et al. (2020),Ball et al. (2021),Yfantis et al. (2024)}. More intriguingly, the light curves of hot-spots present a promising method for detecting chaos, as their overall trend often encodes information about the orbital dynamics of the hot-spot \cite{Huang et al. (2024)}.

In this letter, we employ numerical integration schemes in conjunction with chaotic indicators to obtain chaotic orbits of hot-spots in Kerr spacetime with an external asymptotically uniform electromagnetic field. Using the ray-tracing method, we then simulate the light curves of chaotic hot-spots, thereby establishing the first viable observational window for detecting chaos in celestial dynamical systems.

\emph{Spacetime and orbits}---In Boyer–Lindquist coordinates $x^{\mu}=(t,r,\theta,\varphi)$, the dimensionless covariant metric tensor $g_{\mu\nu}$ of the Kerr spacetime is given by:
\begin{equation}
g_{\mu\nu}=
\begin{pmatrix}\label{1}
-\left(1-\frac{2r}{\Sigma}\right) & 0 & 0 & -\frac{2ar\sin^{2}\theta}{\Sigma} \\
0 & \frac{\Sigma}{\Delta} & 0 & 0 \\
0 & 0 & \Sigma & 0 \\
-\frac{2ar\sin^{2}\theta}{\Sigma} & 0 & 0 & \left(\rho^{2}+\frac{2ra^{2}}{\Sigma}\sin^{2}\theta\right)\sin^{2}\theta
\end{pmatrix},
\end{equation}
where $\Delta=r^{2}-2r+a^{2}$, $\Sigma = r^{2}+a^{2}\cos^{2}\theta$, and $a$ denotes the spin parameter of the black hole. 

Due to the presence of two Killing vector fields in Kerr spacetime, $\xi^{\mu}_{(t)} = (1,0,0,0)$ and $\xi^{\mu}_{(\varphi)} = (0,0,0,1)$, it is convenient to introduce an asymptotically uniform electromagnetic field. The non-zero components of its four-potential are expressed as \cite{Wald (1974)}:
\begin{eqnarray}
A_{t} = -aB\left[1+\frac{r}{\Sigma}\left(\sin^{2}\theta-2\right)\right], \label{2} \\
A_{\varphi} = B\sin^{2}\theta\left[\frac{r^{2}+a^{2}}{2}+\frac{a^{2}r}{\Sigma}\left(\sin^{2}\theta-2\right)\right], \label{3}
\end{eqnarray}
where $B$ denotes the strength of the electromagnetic field. In Kerr spacetime augmented by the electromagnetic field, the motion of a charged hot-spot is governed by the super-Hamiltonian:
\begin{equation}\label{4}
\mathscr{H} = \frac{1}{2}g^{\mu\nu}\left(p_{\mu}-qA_{\mu}\right)\left(p_{\nu}-qA_{\nu}\right),
\end{equation}
where $q$ is the charge of the hot-spot, $p_{\mu}=g_{\mu\nu}\dot{x}^{\mu}$ denotes the conjugate momentum with $\dot{x}^{\mu}$ being the particle’s four-velocity, and $g^{\mu\nu}$ is the contravariant metric tensor.

Since the super-Hamiltonian does not explicitly depend on the coordinate time $t$ or the azimuthal angle $\varphi$, the motion of the hot-spot admits two conserved quantities: the specific energy $E=-p_{t}$ and the specific angular momentum $L=p_{\varphi}$. By employing Hamilton's canonical equations, $\dot{x}^{\mu}=\partial\mathscr{H}/\partial p_{\mu}$ and $\dot{p_{\mu}}=-\partial\mathscr{H}/\partial x_{\mu}$, the trajectory of the hot-spot can be numerically integrated.
\begin{table}
\caption{Parameters $a$, $qB$, $E$, $L$, and orbital states for 6 orbits. All orbits share the same initial generalized coordinates: $t=0$, $r=10.5$, $\theta=\pi/2$, $\varphi=0$; the initial conjugate momentum is set as $p_{r}=0$, and $p_{\theta}$ is derived from the Hamiltonian constraint $\mathscr{H}=-1/2$.}
\label{T1}
\begin{ruledtabular} 
\begin{tabular}{lccccc} 
\toprule
Orbit & $a$ & $qB$ & $E$ & $L$ & State \\
\midrule[1pt]
1 & 0.5 & 0.01 & 0.94 & 3 & Regular \\
2 & 0.9985 & 0.05 & 0.985 & 4.5 & Regular \\
3 & 0.9985 & 0.08 & 0.985 & 4.5 & Regular \\
4 & 0.5 & 0.03 & 0.97 & 4.5 & Chaotic \\
5 & 0.5 & 0.05 & 0.985 & 4.5 & Chaotic \\
6 & 0.3 & 0.1 & 0.96 & 4 & Chaotic \\
\bottomrule
\end{tabular}
\end{ruledtabular}
\end{table}

After obtaining the orbit of the hot-spot, we apply the two-particle method to compute the maximum Lyapunov exponent $\lambda$ \cite{Wu and Huang (2003)} and the Fast Lyapunov Indicator (FLI) \cite{Wu et al. (2006)}to characterize the dynamical nature of the orbit. These quantities are defined as $\lambda =\lim_{t\rightarrow\infty}(1/t)\ln(d_{t}/d_{0})$ and $\textrm{FLI} = \log_{10}\left(d_{t}/d_{0}\right)$, where $d_{0}$ and $d_{t}$ denote the phase-space distance between two neighboring trajectories at the initial time and at time $t$, respectively. Here, we set $d_{0}=10^{-8}$.

We selected $6$ orbits with different initial conditions, and the corresponding parameters are summarized in Tab. I. Figs. 1 and 2 present the trajectories (left column), dynamical parameters (middle column), and chaotic indicators (right column) for orbits 1--3 and orbits 4--6, respectively. For the regular orbits 1--3, the trajectories exhibit clear and coherent patterns. The evolution of the orbital radial coordinate, latitude, and azimuthal angle over time demonstrates periodic or quasi-periodic behavior. Moreover, the maximum Lyapunov exponents approaching zero, along with the linear and slow growth of the FLI, provide robust evidence of their regular nature.

In contrast, the three chaotic orbits shown in Fig. 2 exhibit irregular and unpredictable behavior, with orbital parameters displaying more complex evolutionary patterns and no discernible regularity. Specifically, the maximum Lyapunov exponents of these orbits begin to diverge around $t = 10^{3}$M, eventually converging to a positive, non-zero value as the integration time increases. Meanwhile, the FLI grows exponentially with coordinate time, at a significantly faster rate than the corresponding values in Fig. 1.
\begin{figure*}
\center{
\includegraphics[width=4cm]{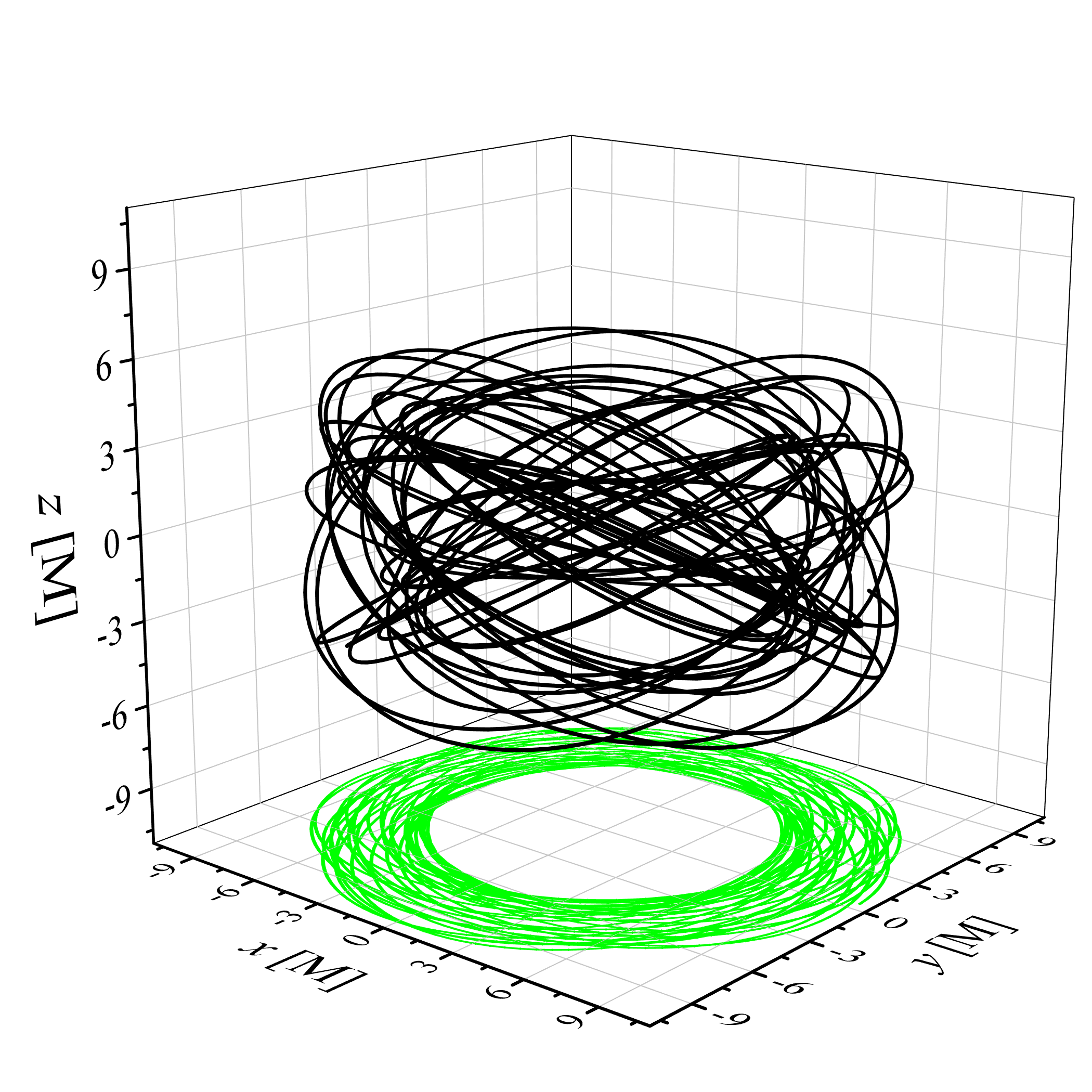}
\includegraphics[width=8.57cm]{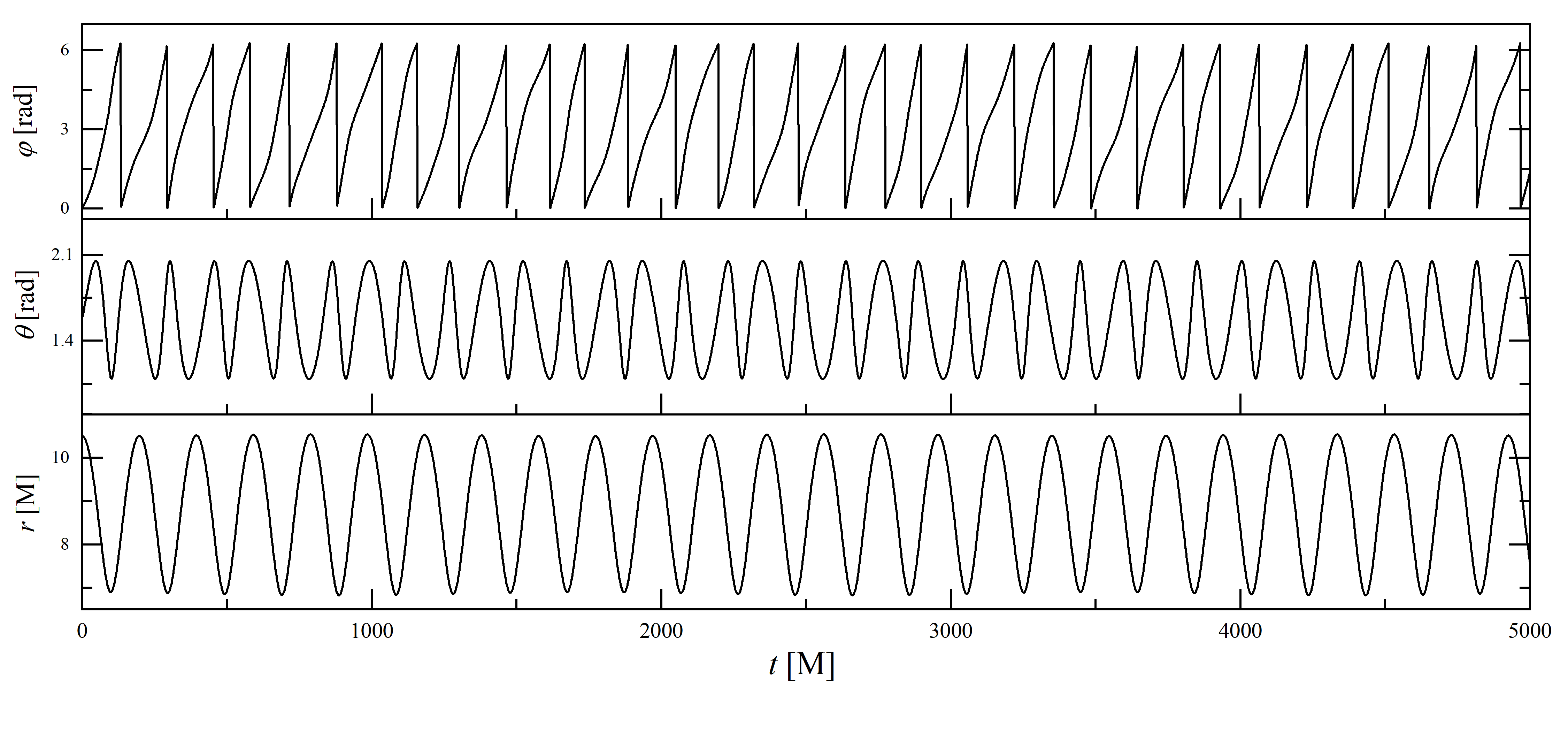}
\includegraphics[width=4cm]{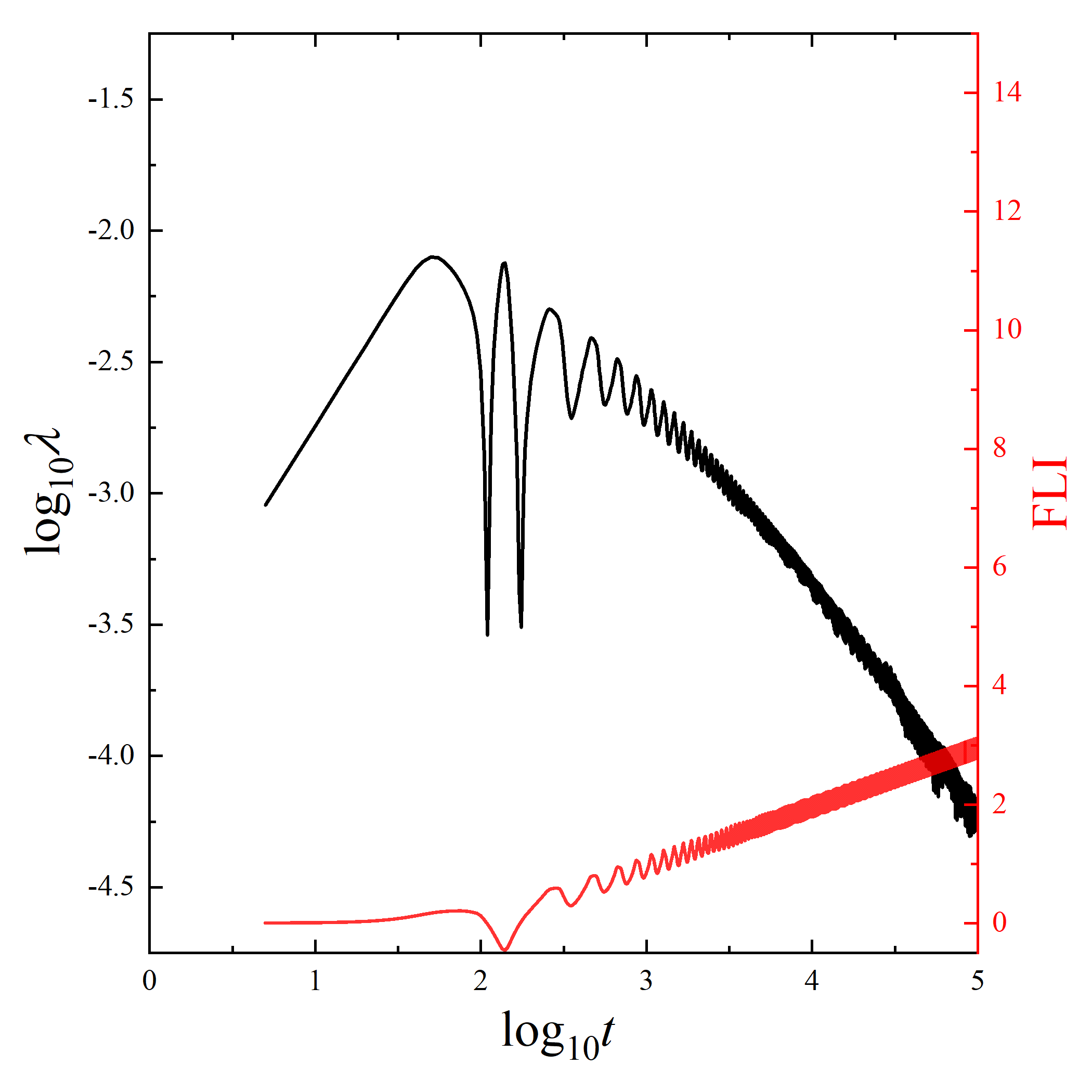}
\includegraphics[width=4cm]{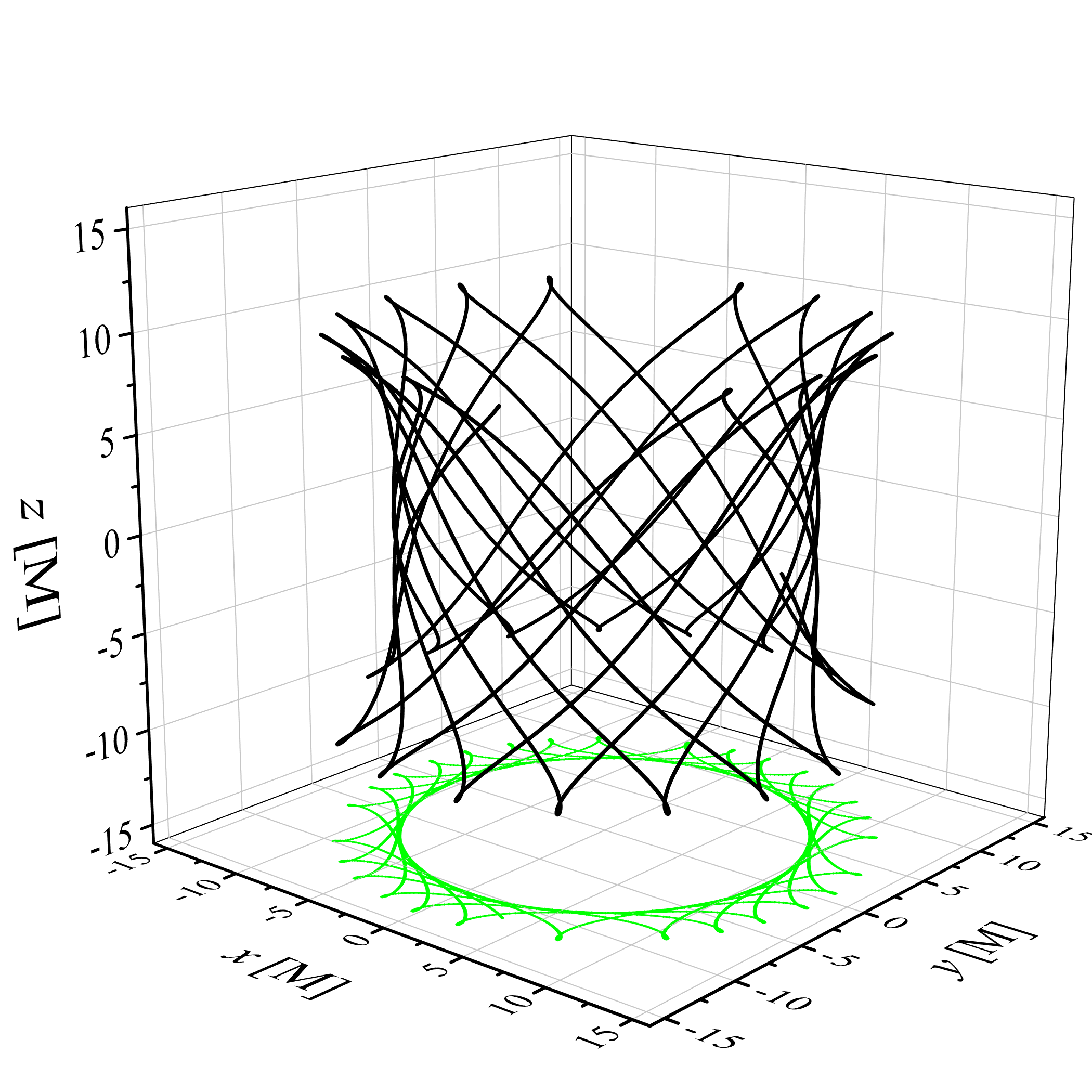}
\includegraphics[width=8.57cm]{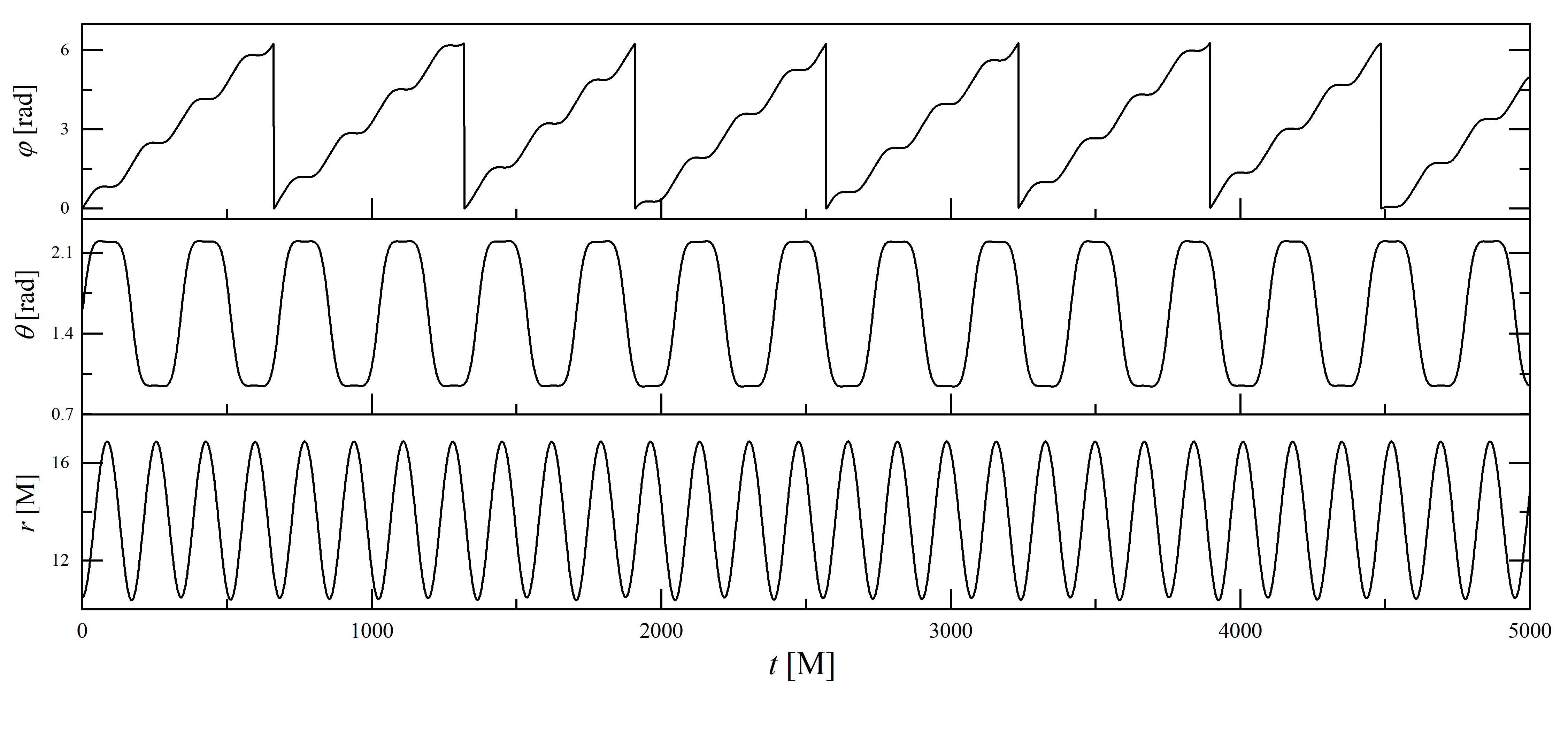}
\includegraphics[width=4cm]{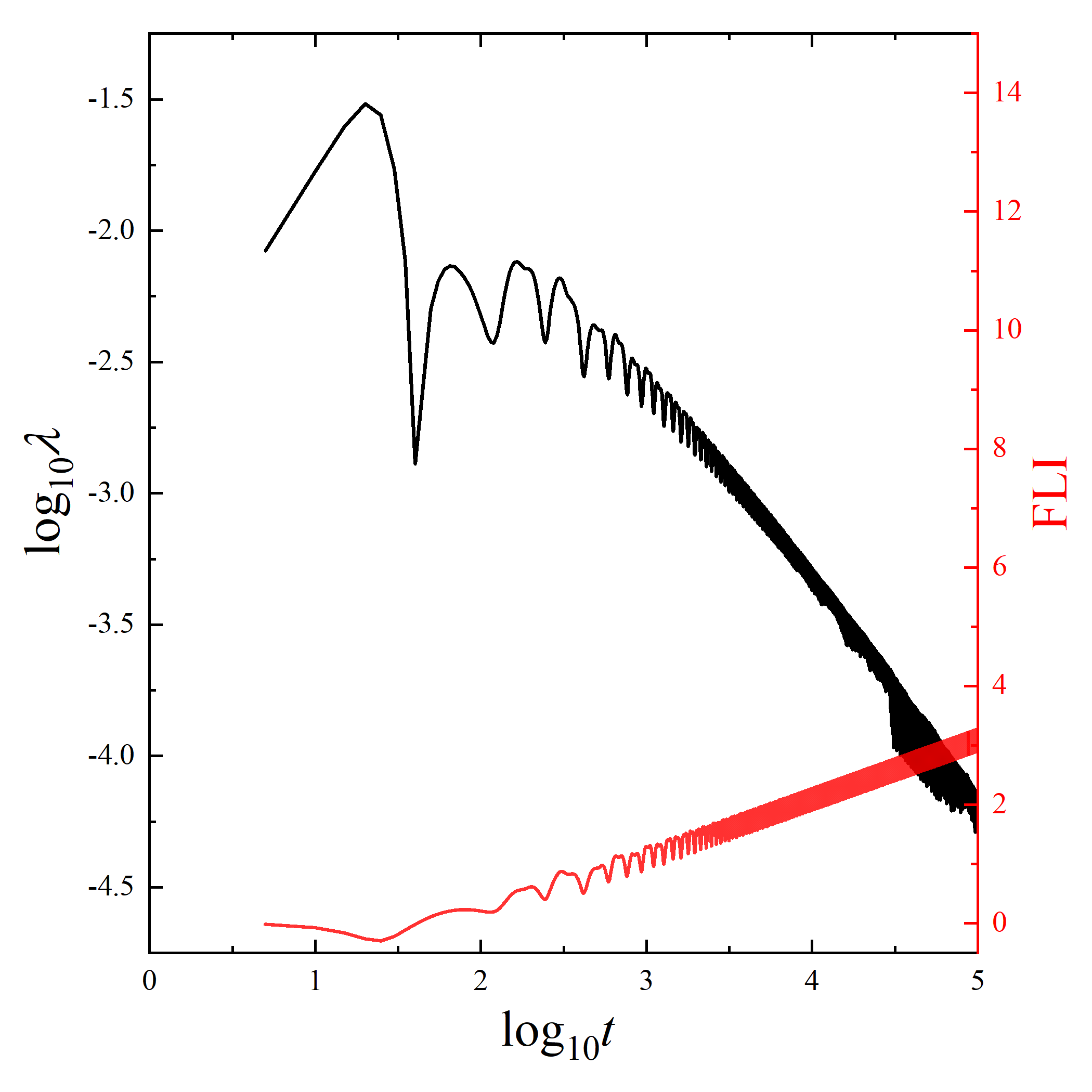}
\includegraphics[width=4cm]{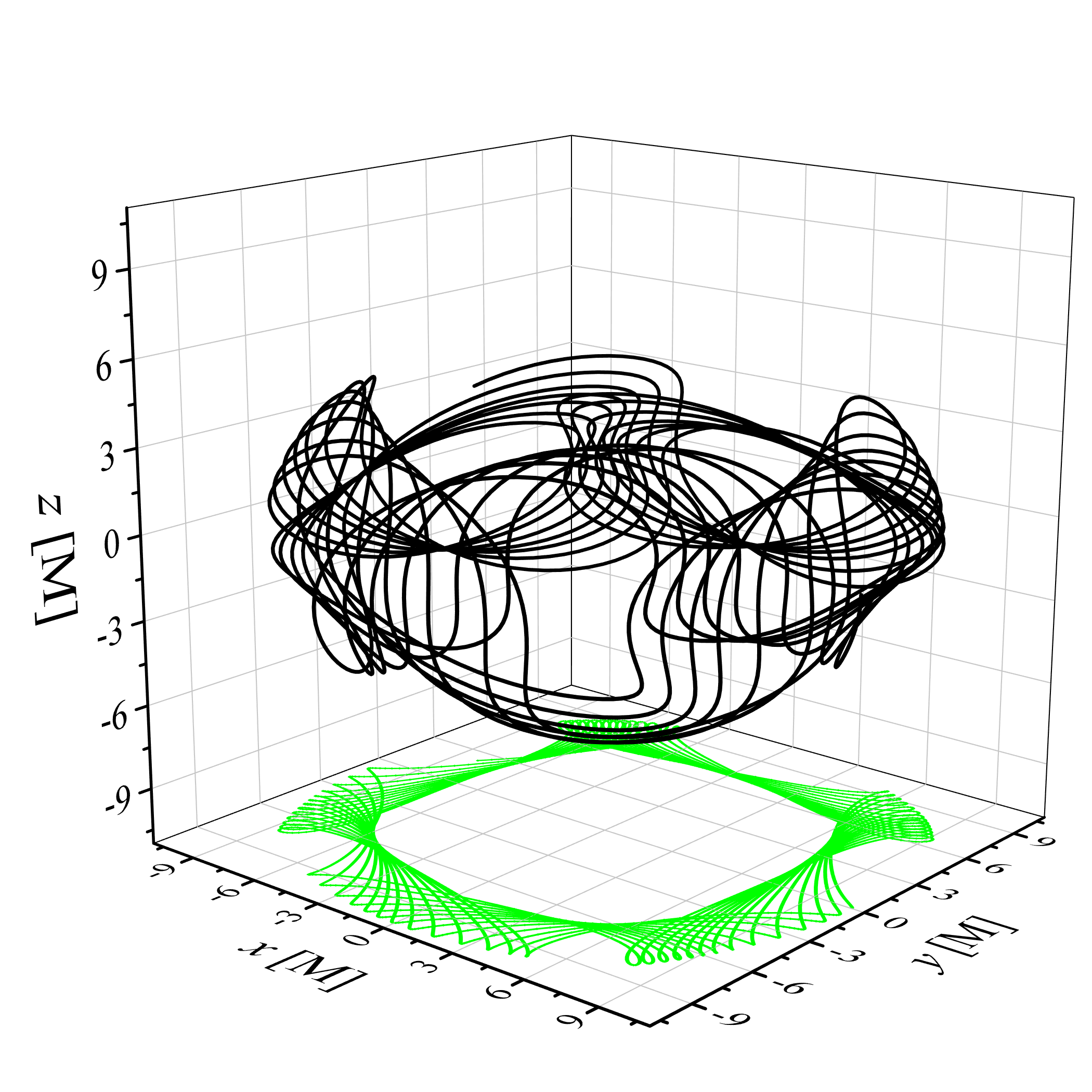}
\includegraphics[width=8.57cm]{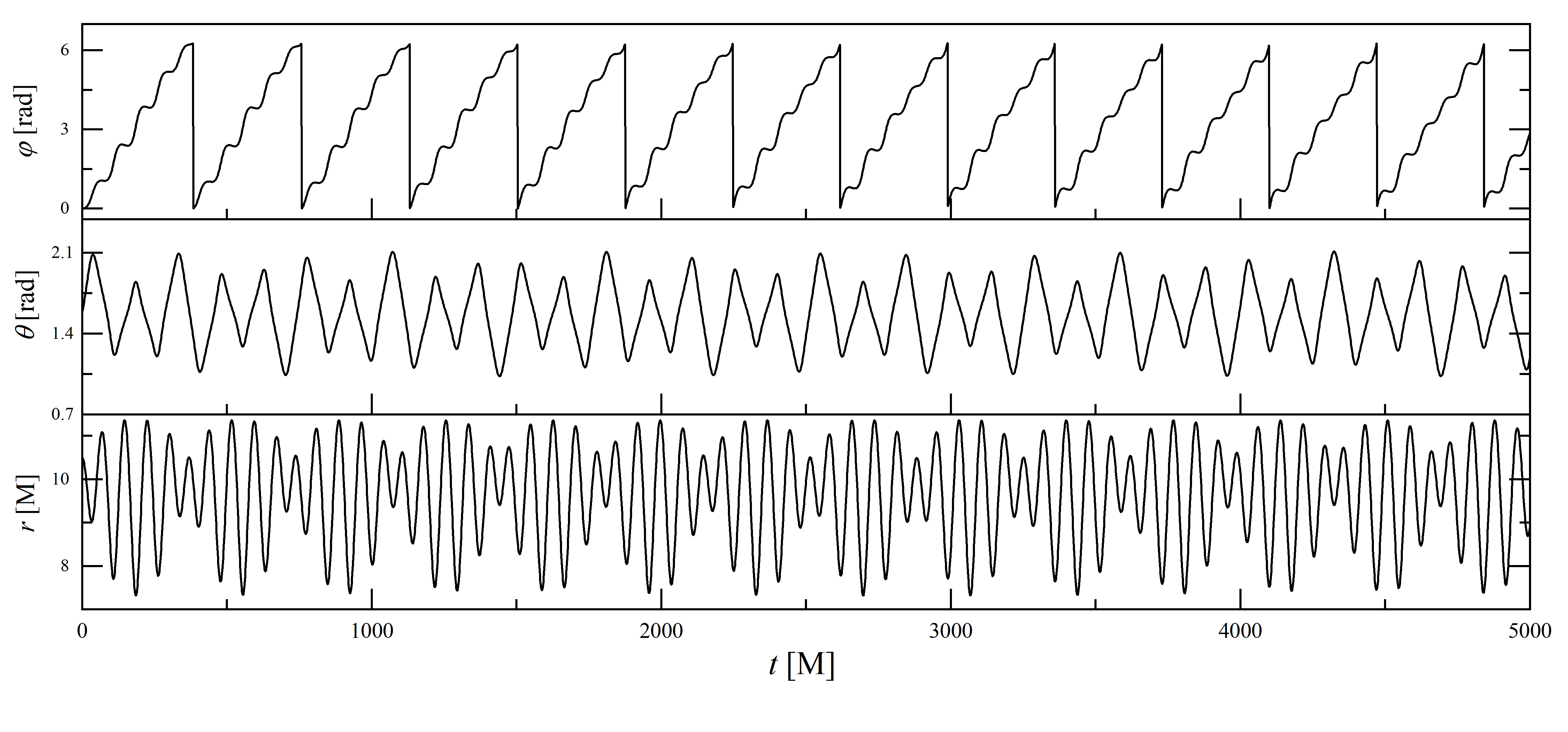}
\includegraphics[width=4cm]{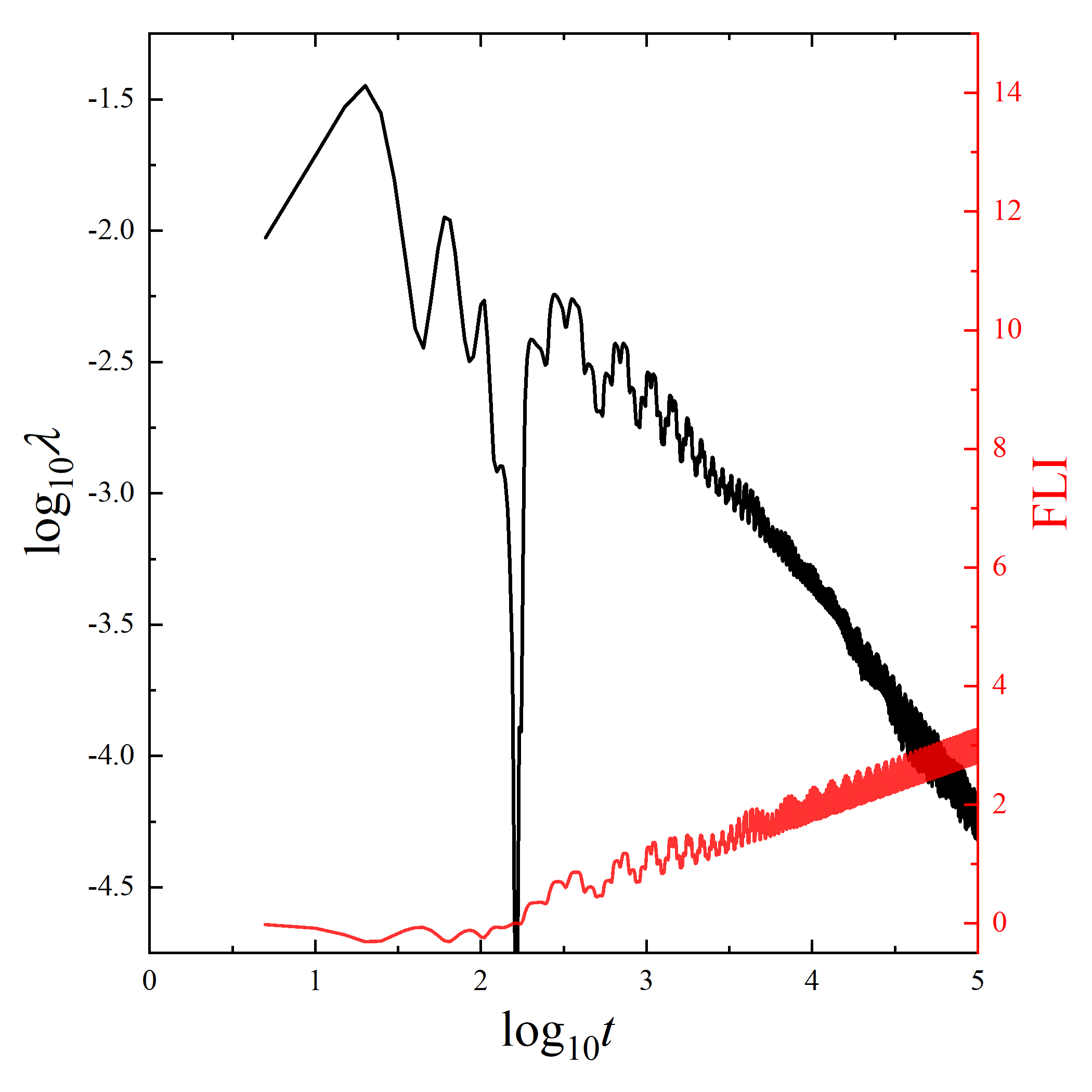}
\caption{(colour online) Trajectories (left column), orbital parameters (middle column), and chaotic indicators (right column) for orbits 1--3 (top to bottom). These orbits exhibit distinct periodic or quasi-periodic patterns in their trajectories. The orbital parameters---such as $\varphi$, $\theta$, and $r$---also display periodic oscillatory behavior. The near-zero maximum Lyapunov exponents (black curves) and the slowly, linearly increasing Fast Lyapunov Indicators (red curves) on the right provide key numerical evidence supporting the regular nature of these orbits.}}\label{fig1}
\end{figure*}

\begin{figure*}
\center{
\includegraphics[width=4cm]{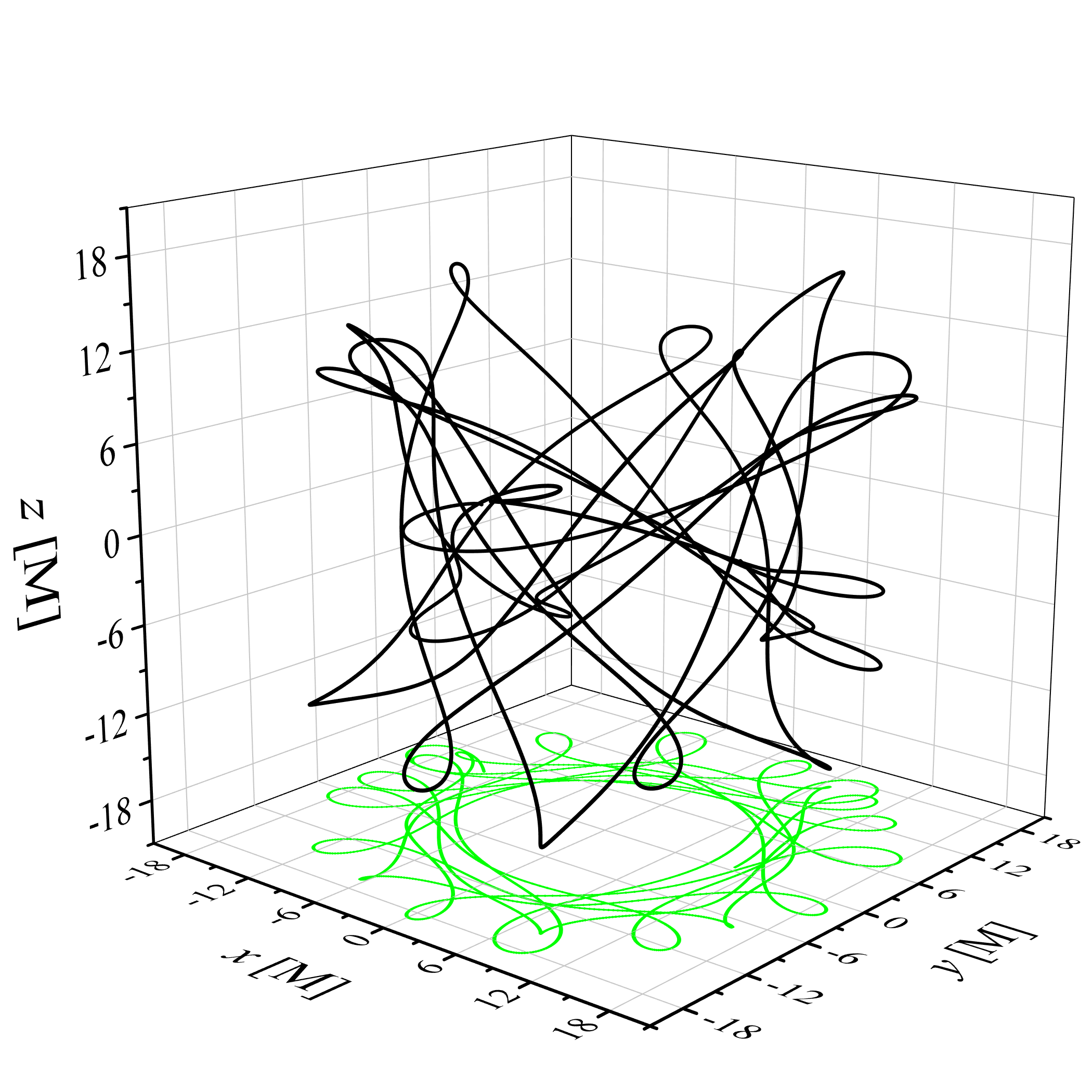}
\includegraphics[width=8.57cm]{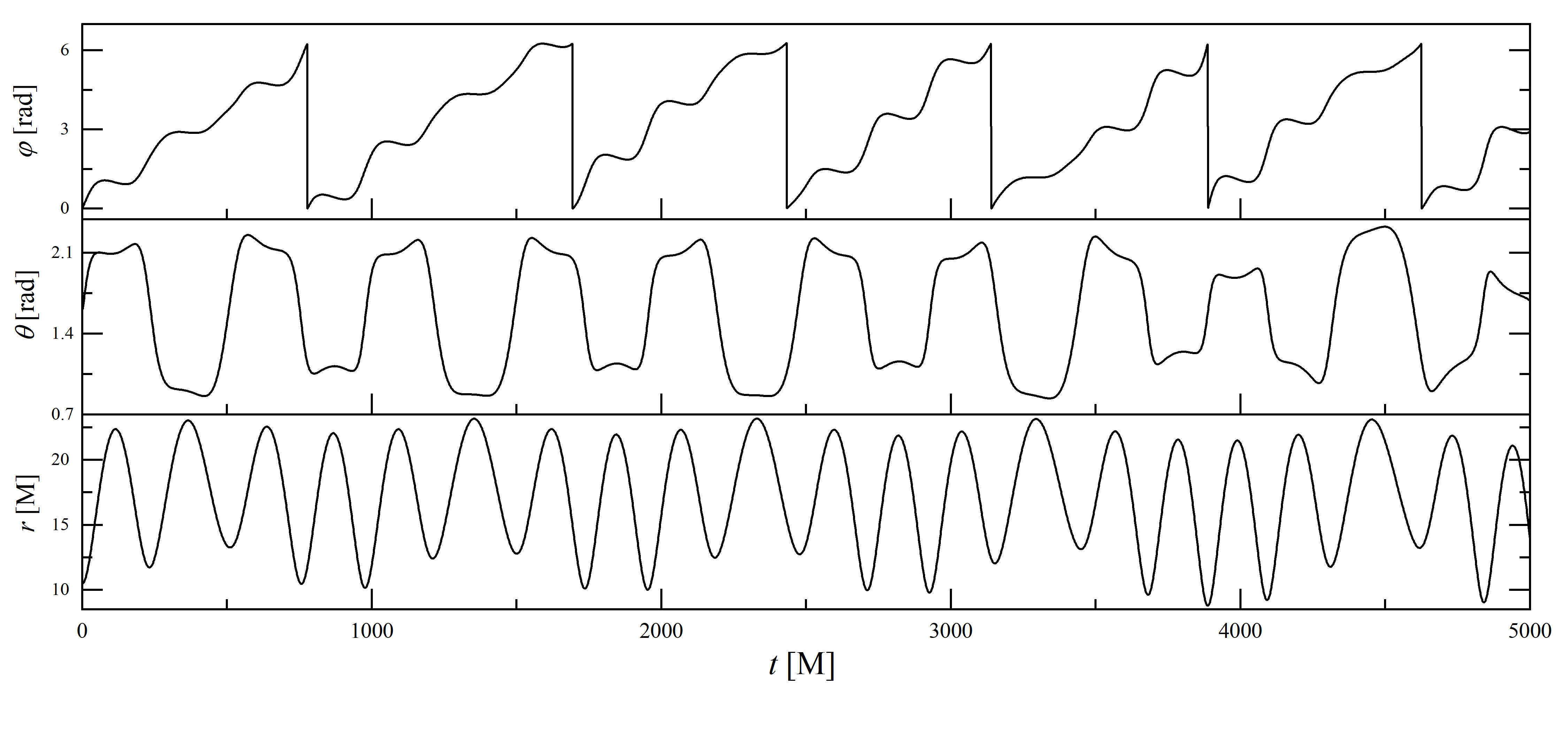}
\includegraphics[width=4cm]{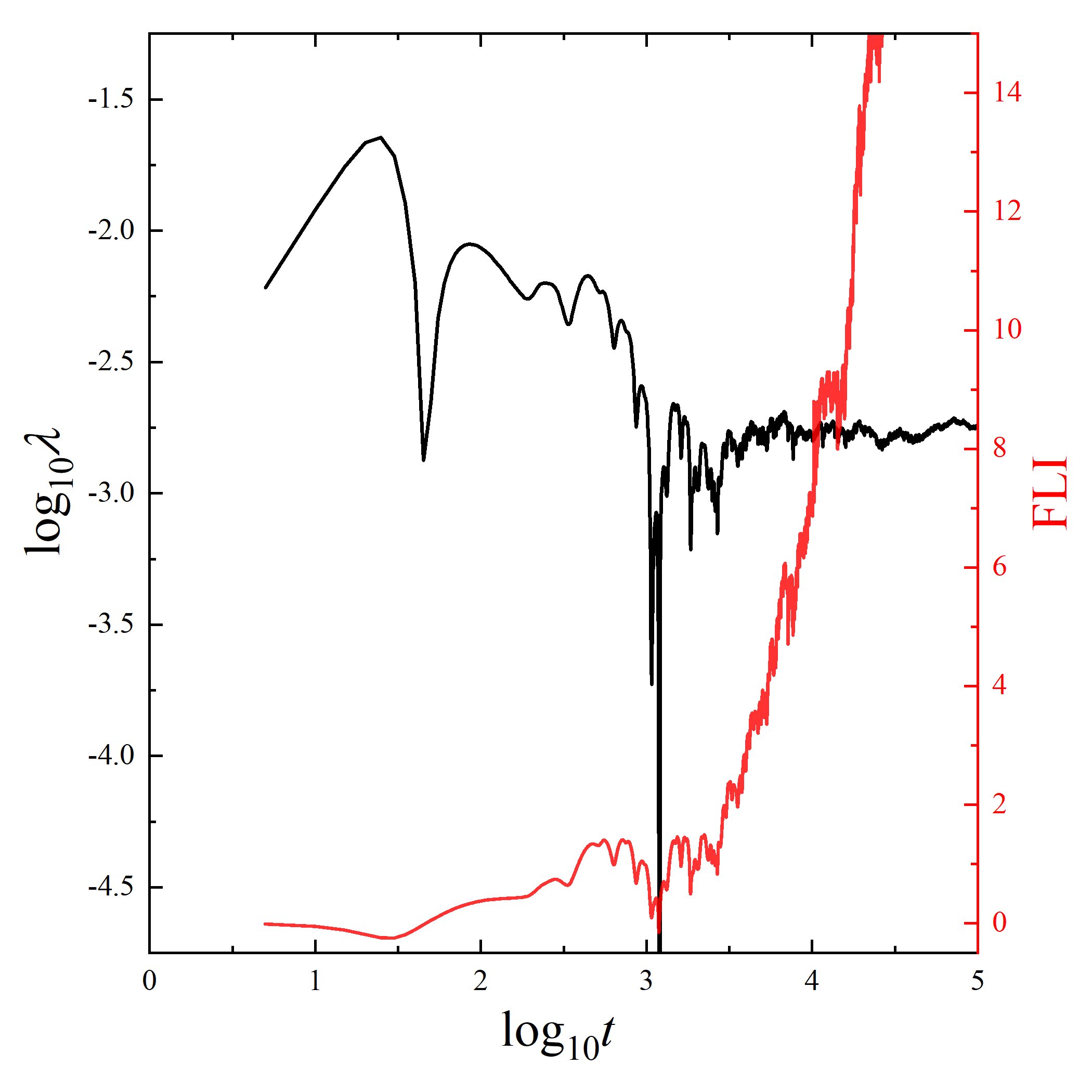}
\includegraphics[width=4cm]{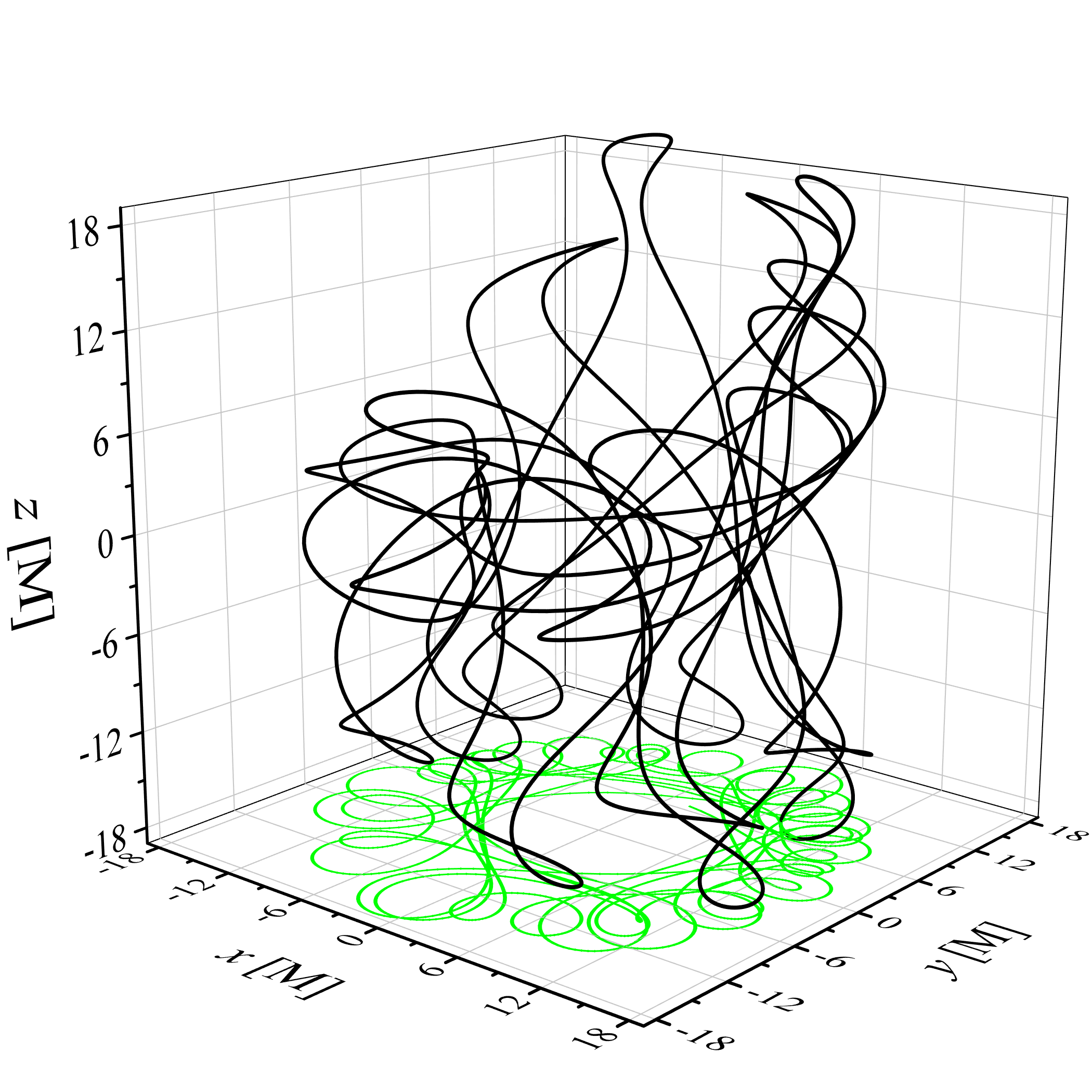}
\includegraphics[width=8.57cm]{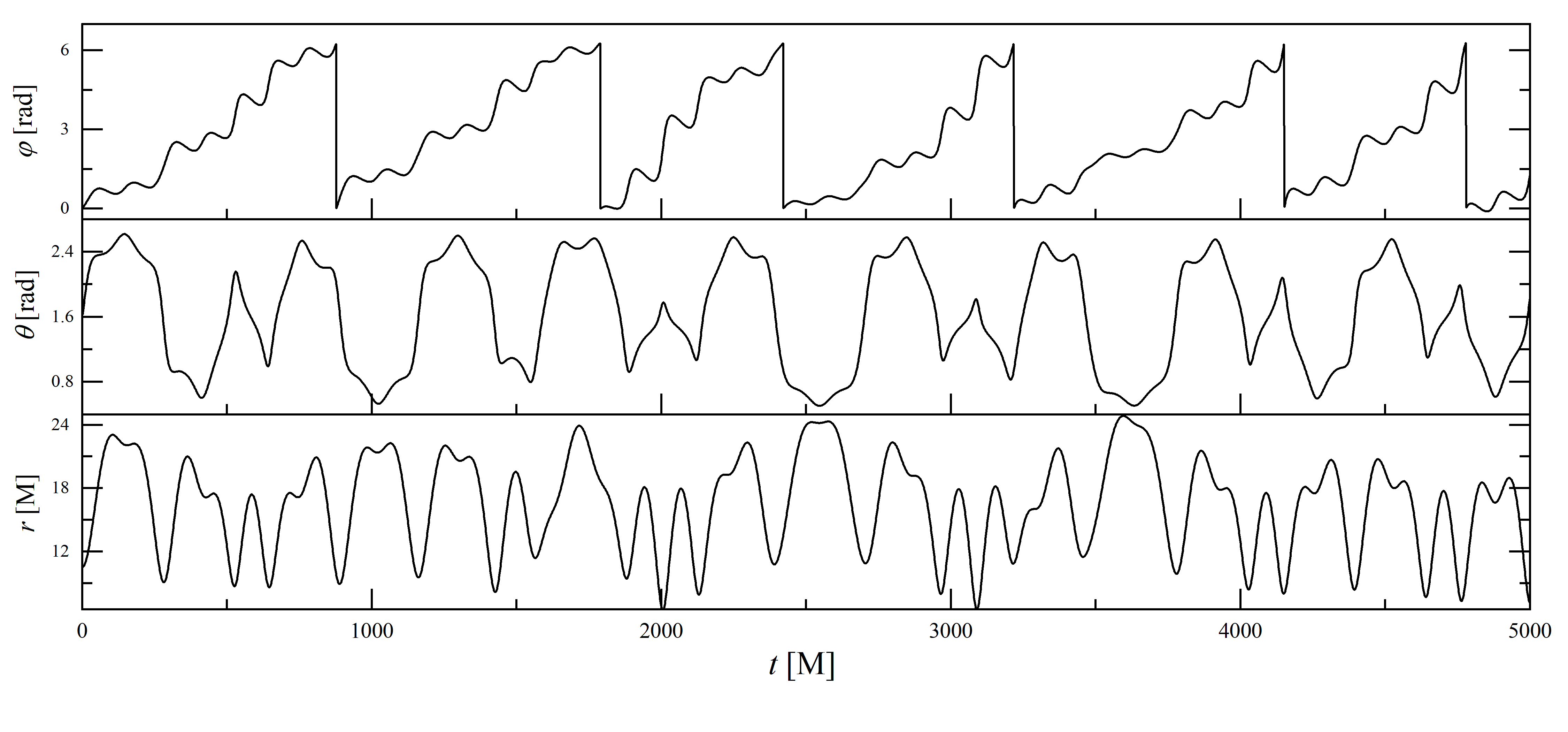}
\includegraphics[width=4cm]{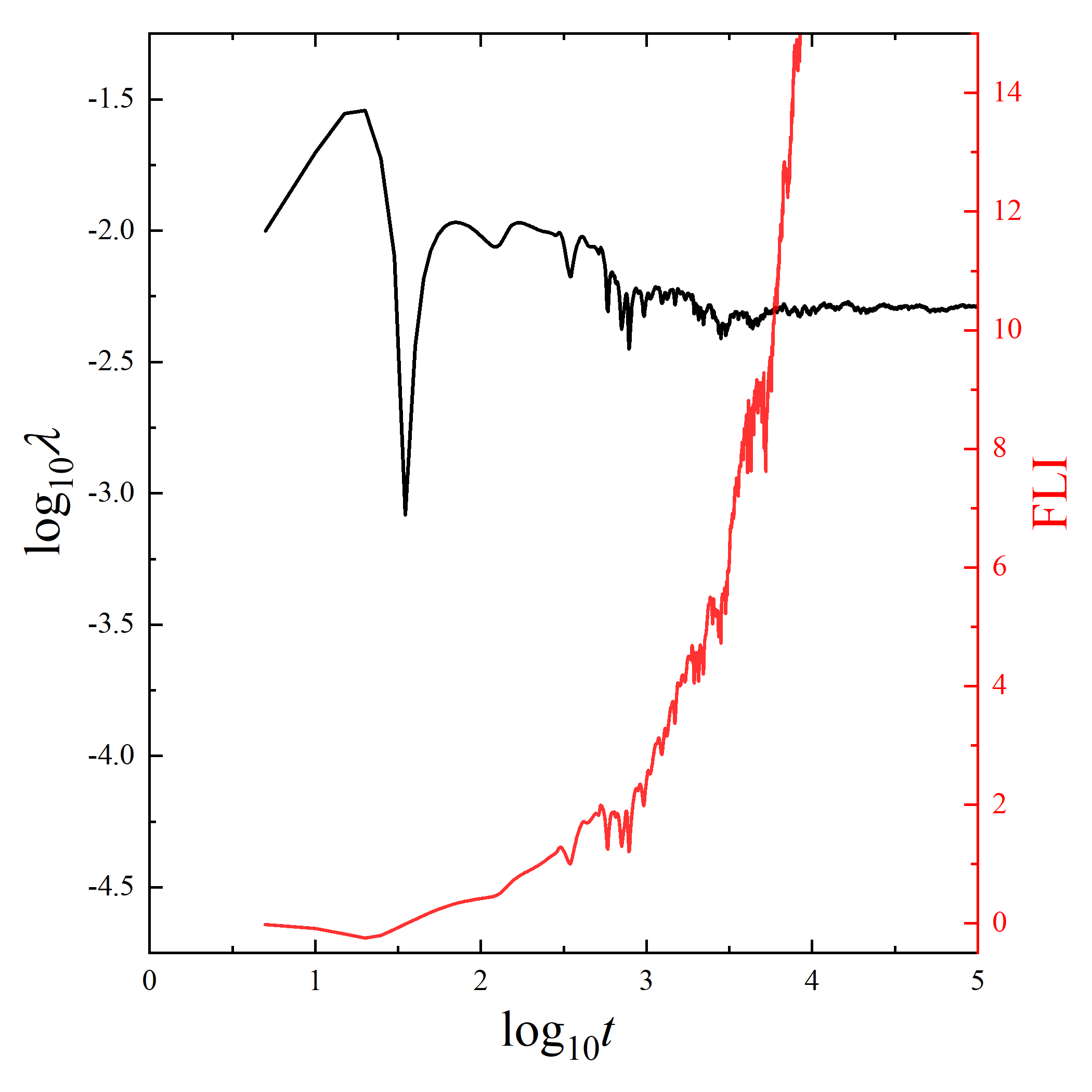}
\includegraphics[width=4cm]{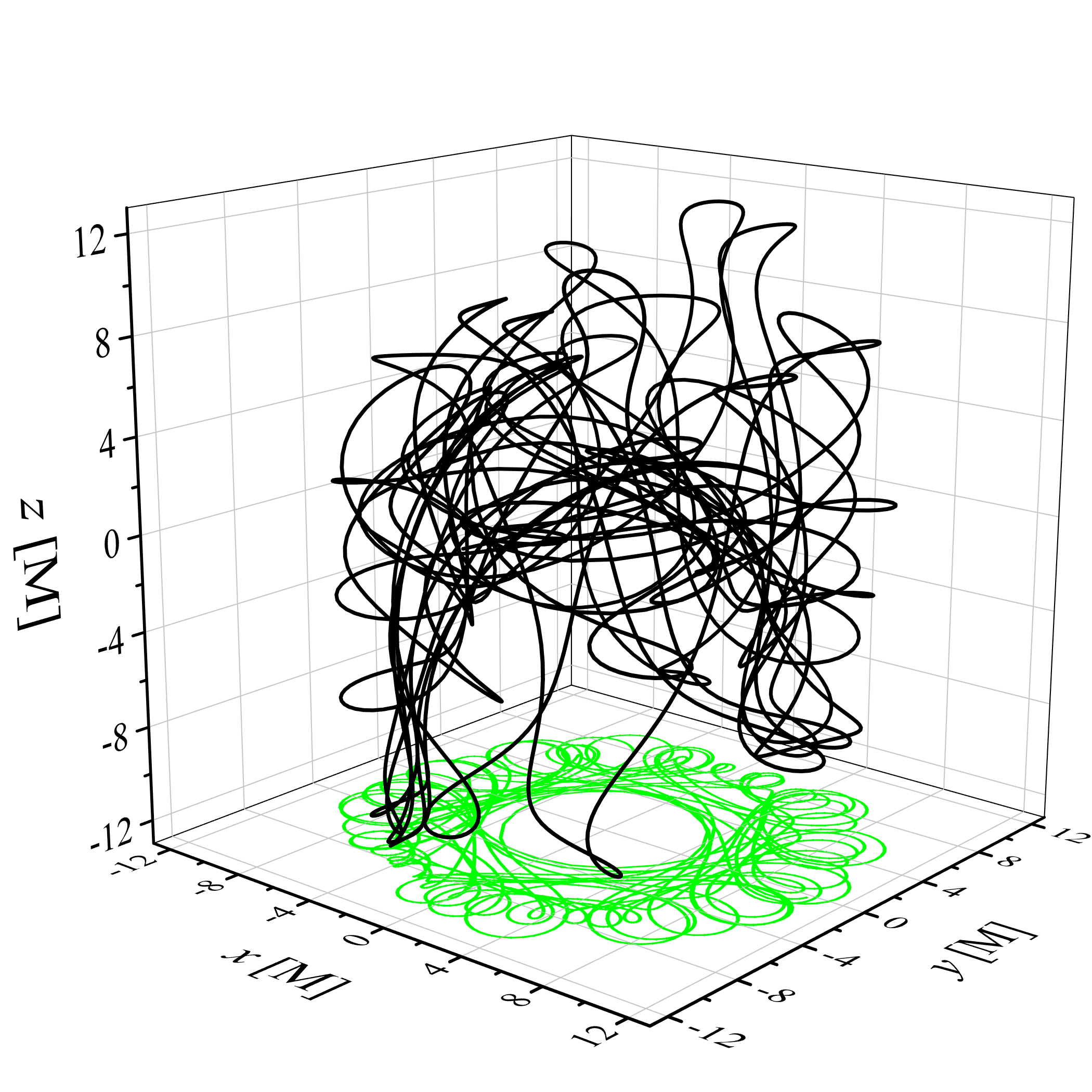}
\includegraphics[width=8.57cm]{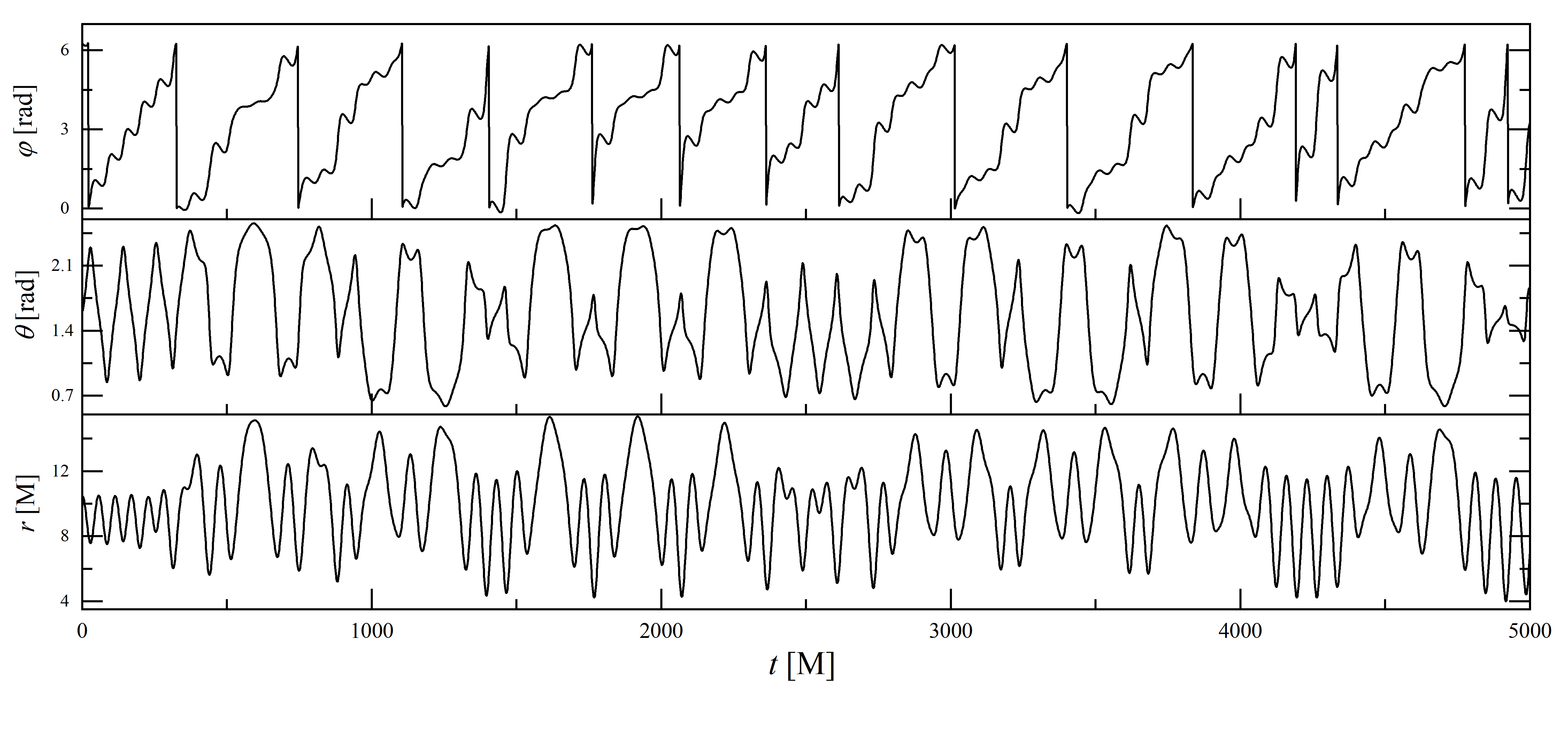}
\includegraphics[width=4cm]{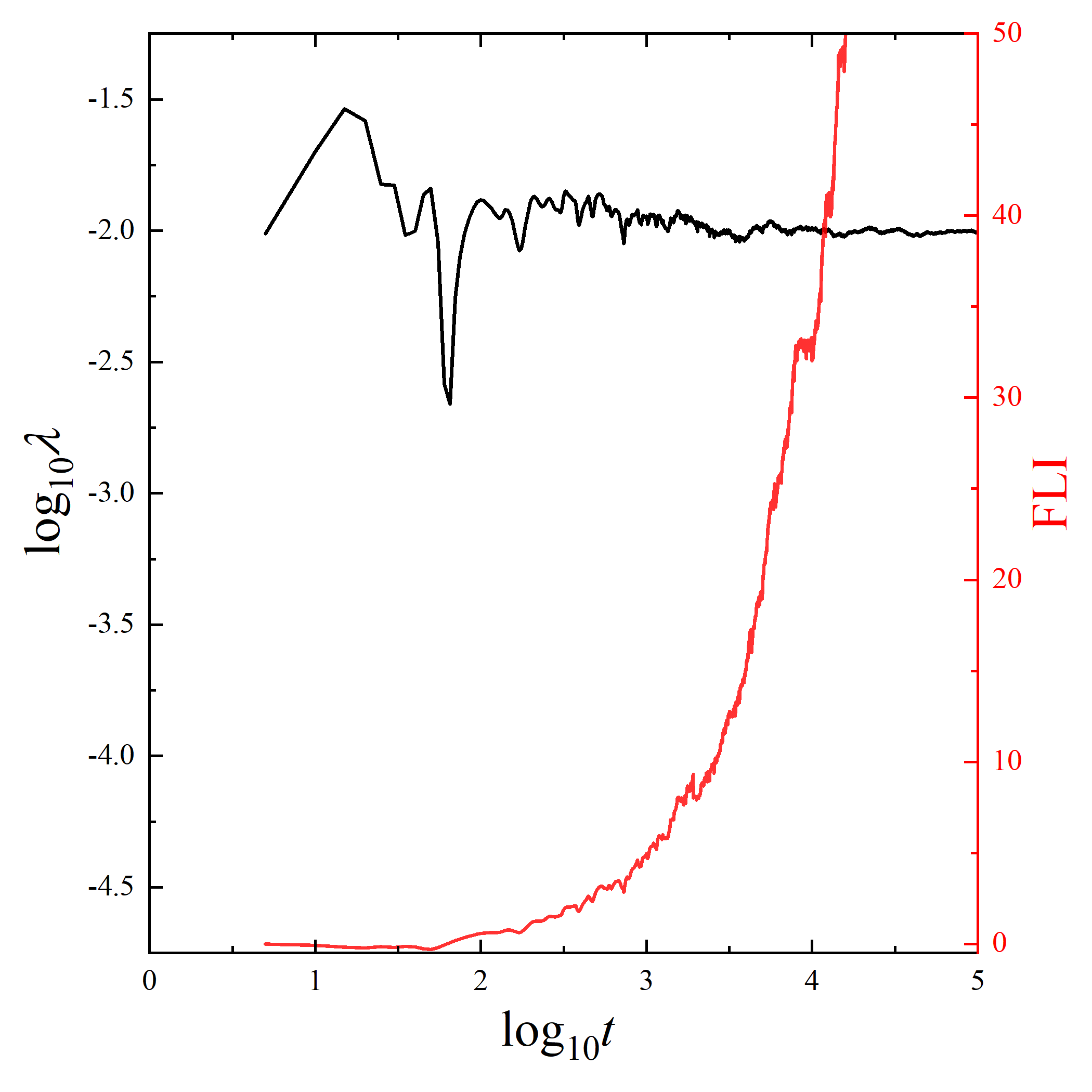}
\caption{(colour online) Similar to Fig. 1, but for chaotic orbits 4--6. When the hot-spot exhibits chaotic motion, the trajectories resemble a tangled bundle with no discernible order. The evolution of the orbital parameters over time further emphasizes their unpredictable nature. More importantly, the maximum Lyapunov exponents of these orbits converge to a positive, non-zero value over time, while the Fast Lyapunov Indicators grow exponentially.}}\label{fig2}
\end{figure*}

\emph{Signatures of chaotic hot-spots in light curves.}---We fix the observer's inclination and azimuthal angles to $17^{\circ}$ and $0^{\circ}$, respectively, and set the field of view and resolution to $x, y \subset [-30, 30]$ M and $6000 \times 6000$ pixels. The light curves for orbits 1--6 are simulated following the ray-tracing method implemented in ODYSSEY \cite{Pu et al. (2016)}. Specifically, light rays are emitted from each grid point in the observer's image plane and traced backward in time to determine whether they intersect the trajectory of the hot-spot. If an intersection occurs, the corresponding ray contributes to the photon count rates at time $t_{\textrm{s}} + \tau$, where $t_{\textrm{s}}$ denotes the hot-spot's coordinate time and $\tau$ is the light travel time from the observer to the hot-spot.

The left panel of Fig. 3 shows the light curves corresponding to the motion of the hot spot along orbits 1 to 6. Among these, the light curves of orbits 1 and 2 exhibit clearly discernible periodic fluctuations, reminiscent of a heartbeat. Their power spectra, obtained via fast Fourier transform, reveal clean, isolated, sharp, and narrow peaks, as shown in the right panel. Although identifying clear periodicity in the light curve of orbit 3 is more challenging, its power spectrum similarly exhibits prominent, sharp, narrow peaks. These results suggest that the regular characteristics of the orbits can be encoded in the light curves.

In contrast, for the chaotic orbits (last three rows in Fig. 3), the light curves resemble a cluttered horizon with no discernible regularity. The range of fluctuations in these curves is clearly much broader than in the regular case. In the frequency domain, their power spectra exhibit continuous, low-amplitude broad peaks resembling noise. We conclude that the light curves of hot-spots can encode information about orbital regularity or chaos, particularly through their power spectra. This provides a solid theoretical foundation for detecting chaotic orbits in curved spacetime. Furthermore, chaotic motion typically arises only when system parameters exceed certain thresholds, implying that chaotic orbits identified through light curves may also offer insights into the properties of curved spacetime.
\begin{figure*}
\center{
\includegraphics[width=8.5cm]{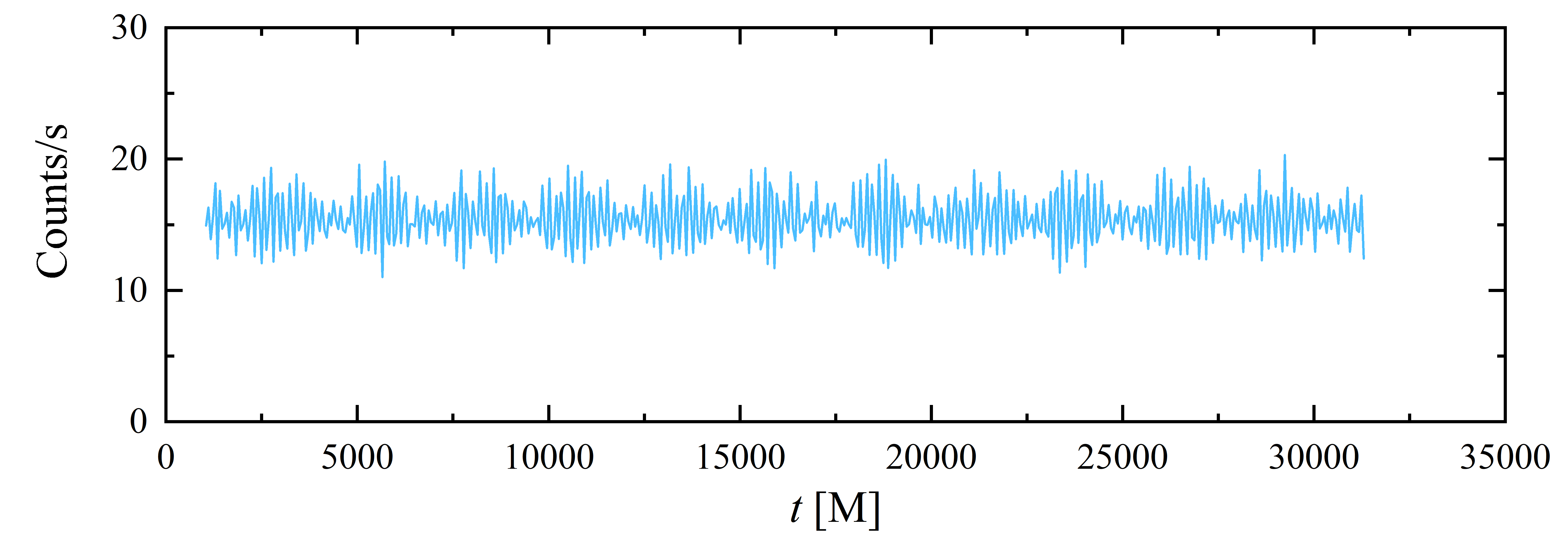}
\includegraphics[width=8.5cm]{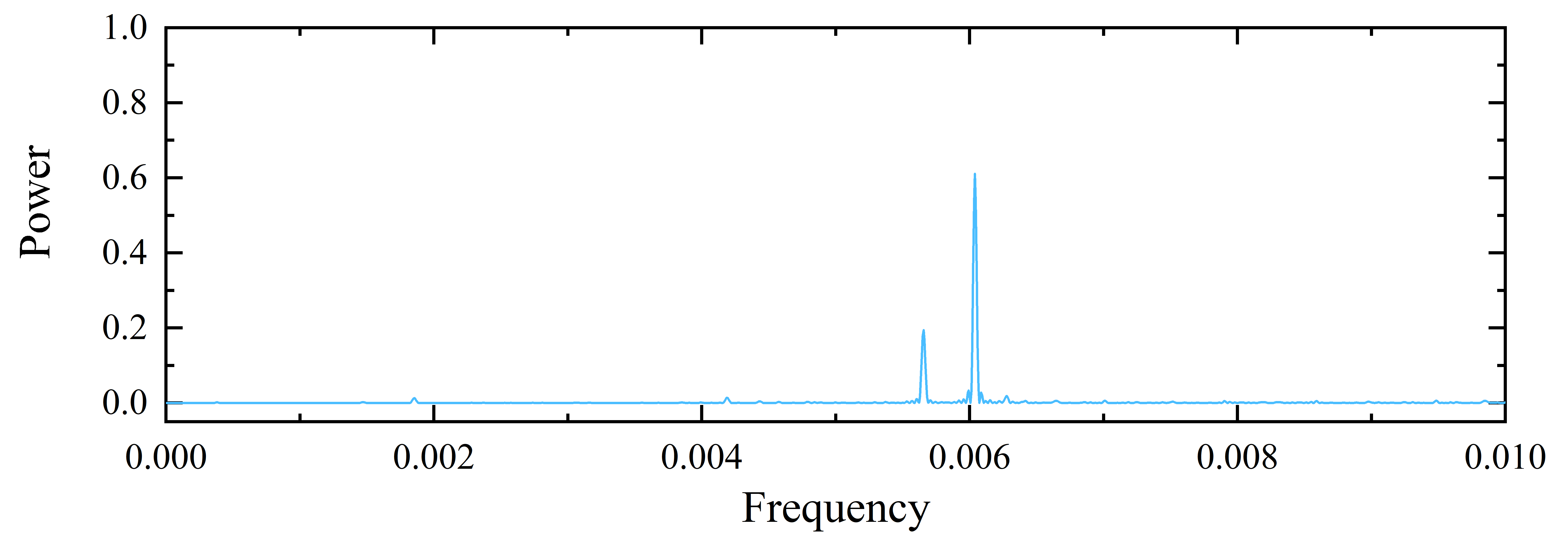}
\includegraphics[width=8.5cm]{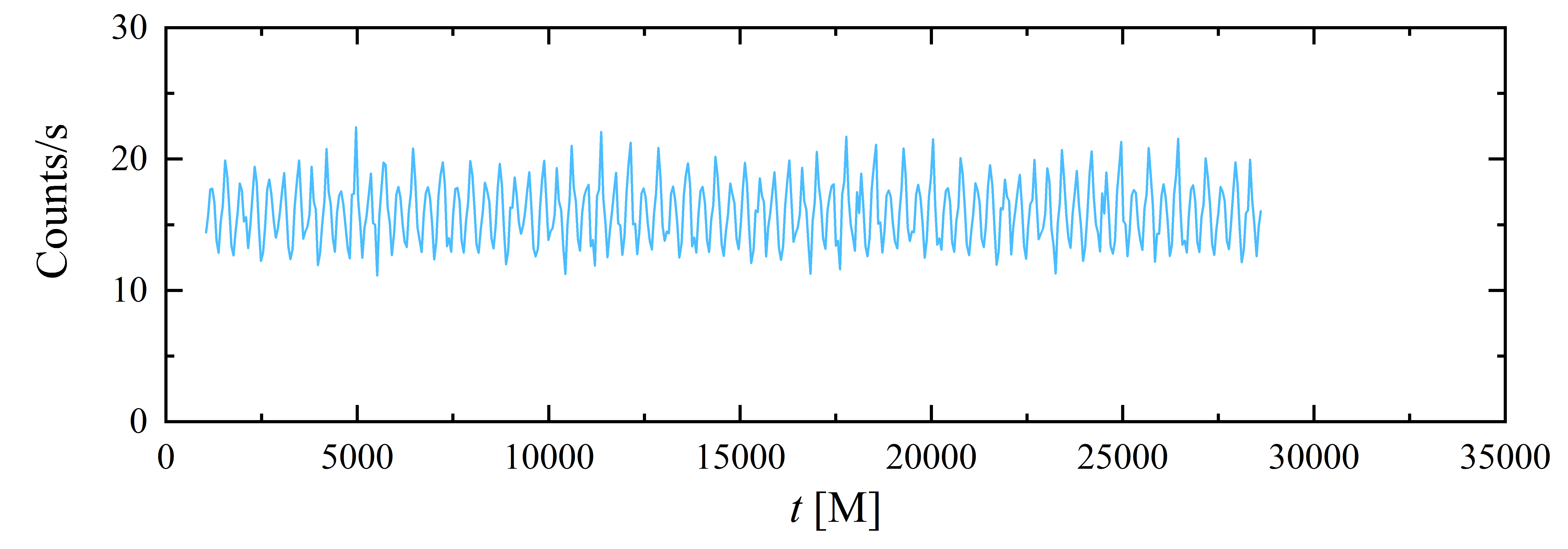}
\includegraphics[width=8.5cm]{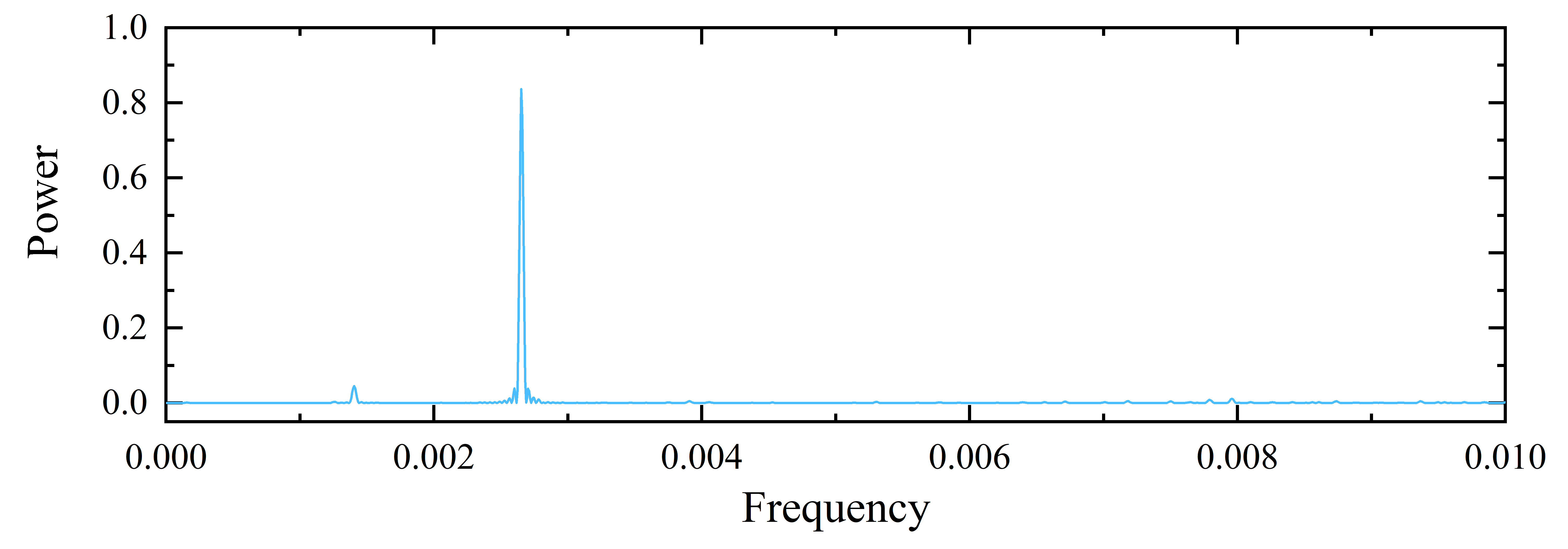}
\includegraphics[width=8.5cm]{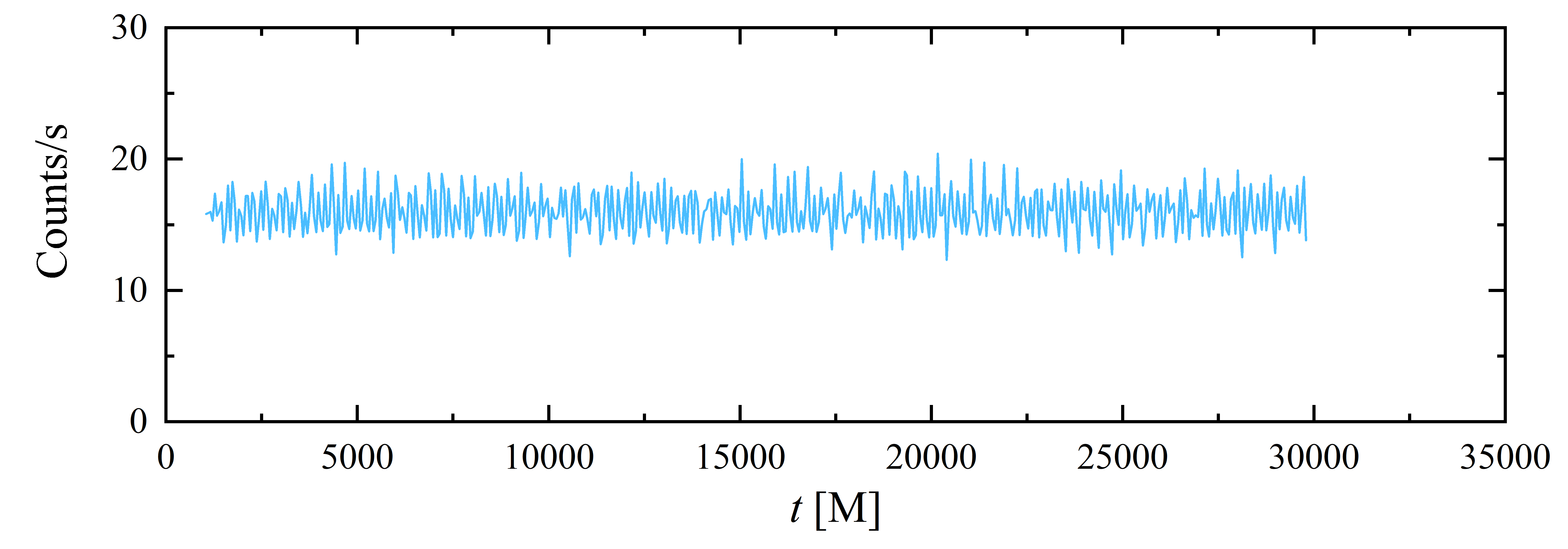}
\includegraphics[width=8.5cm]{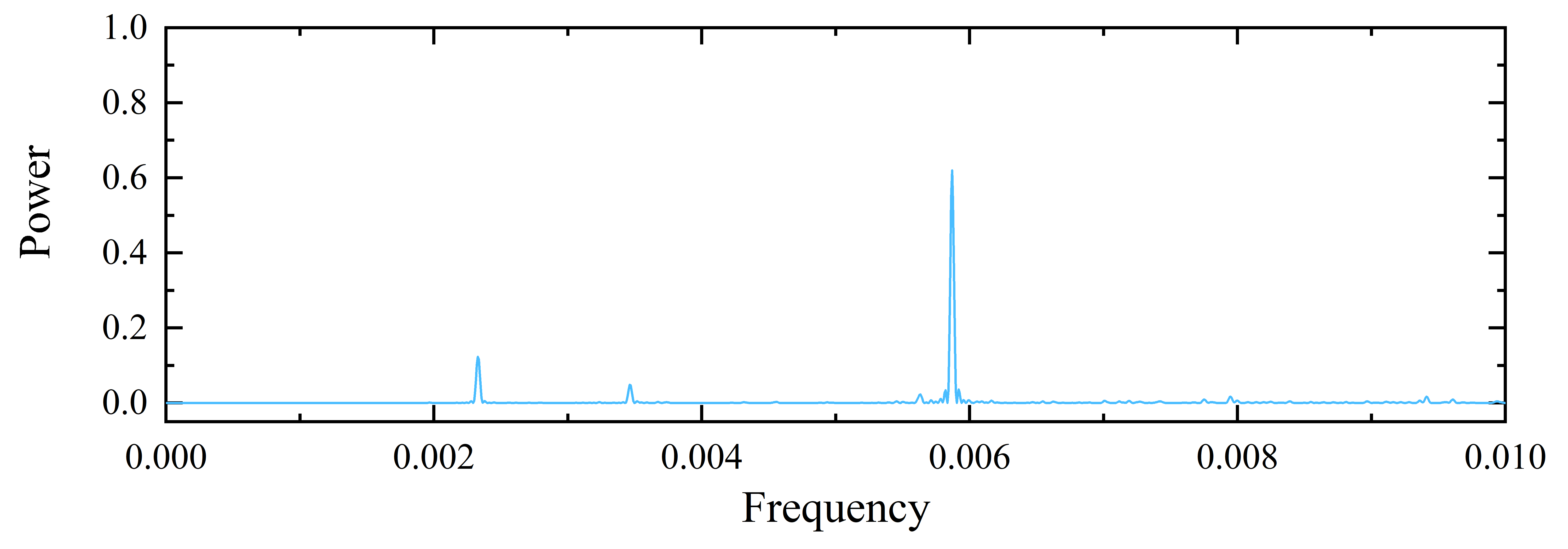}
\includegraphics[width=8.5cm]{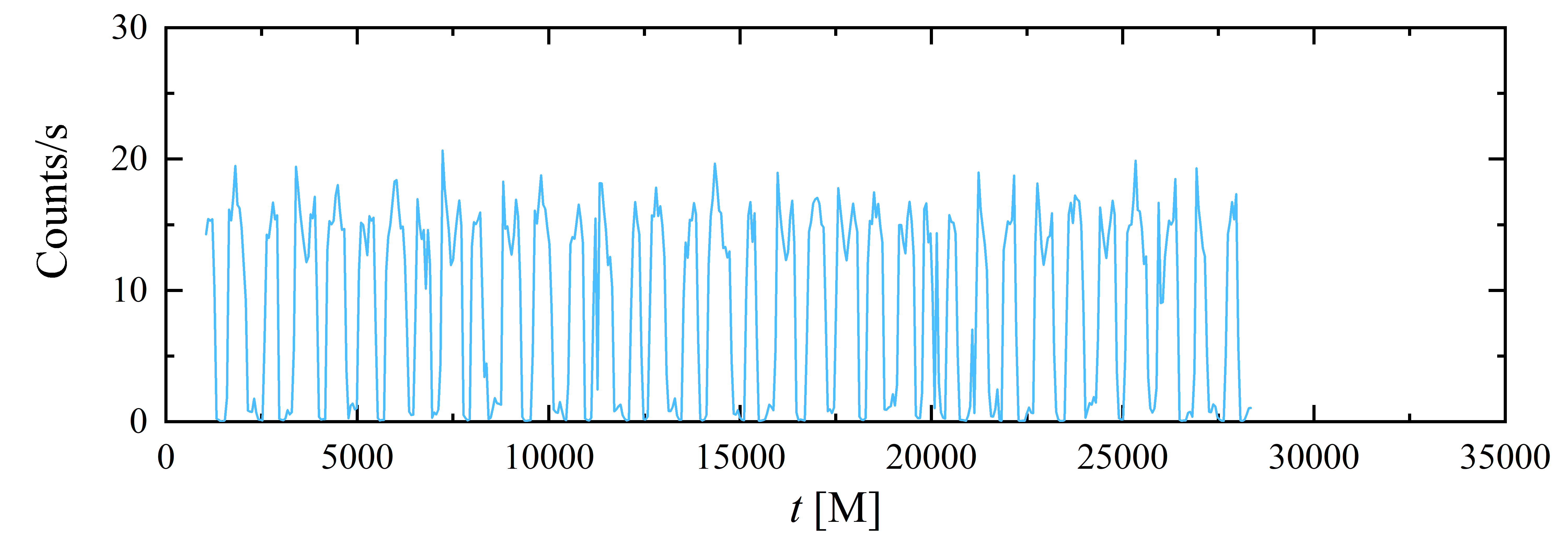}
\includegraphics[width=8.5cm]{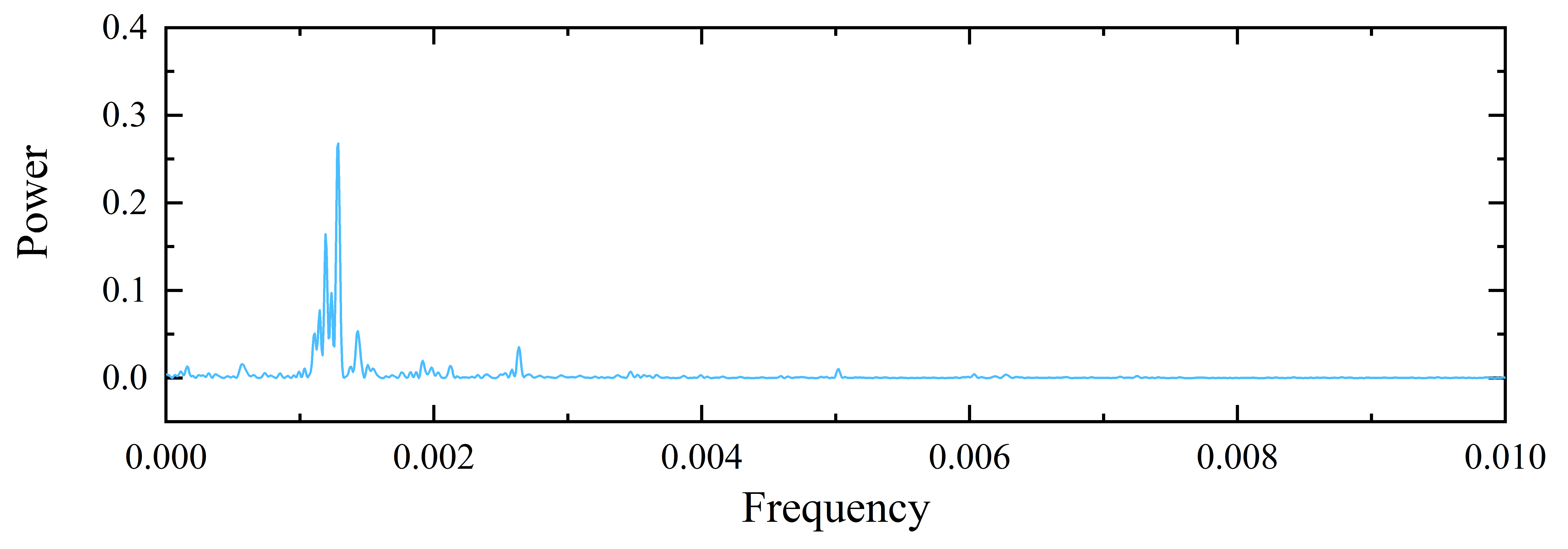}
\includegraphics[width=8.5cm]{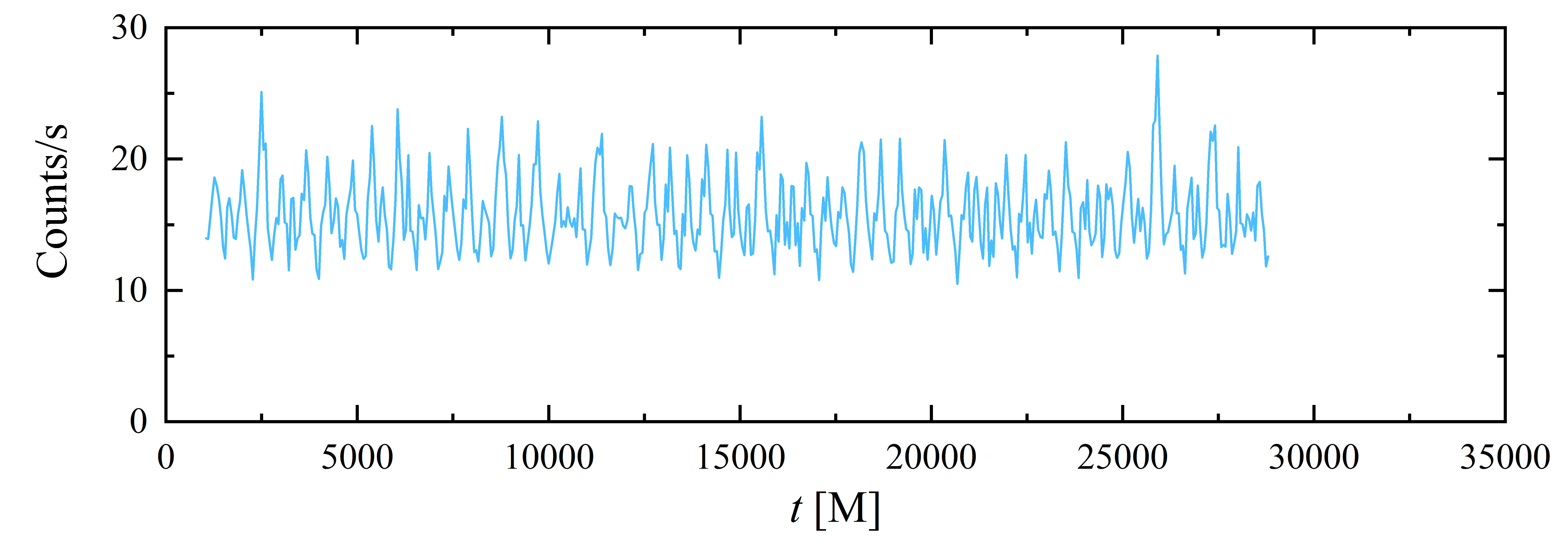}
\includegraphics[width=8.5cm]{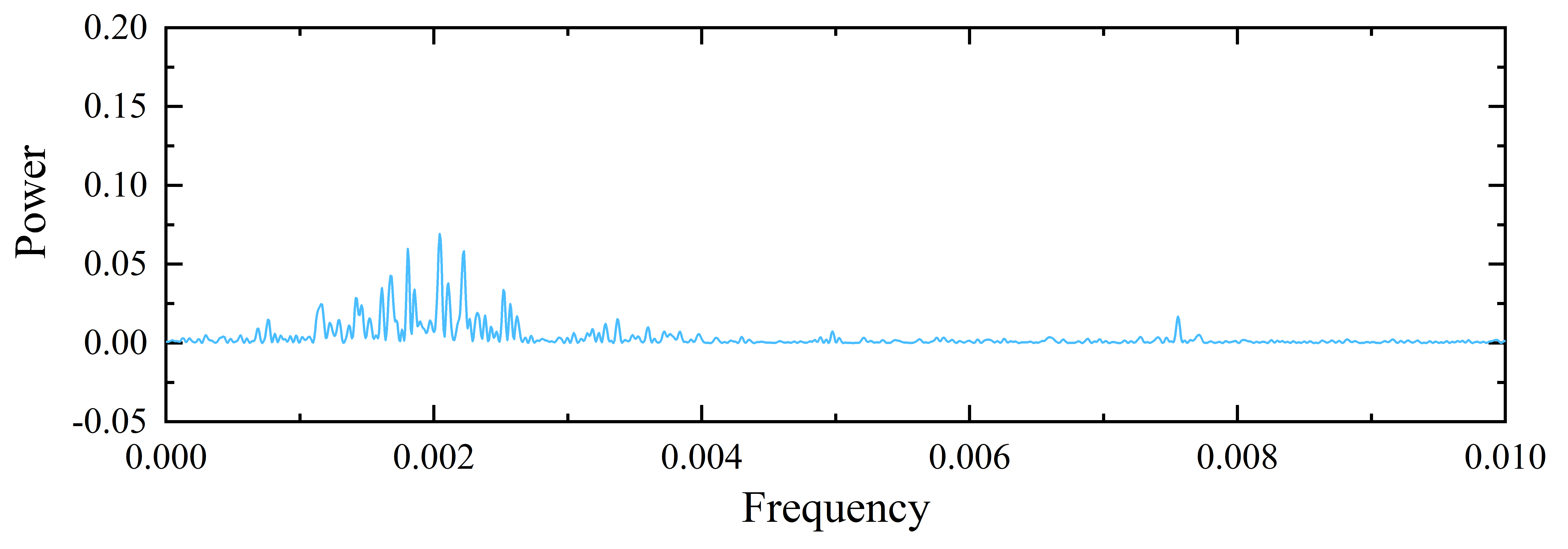}
\includegraphics[width=8.5cm]{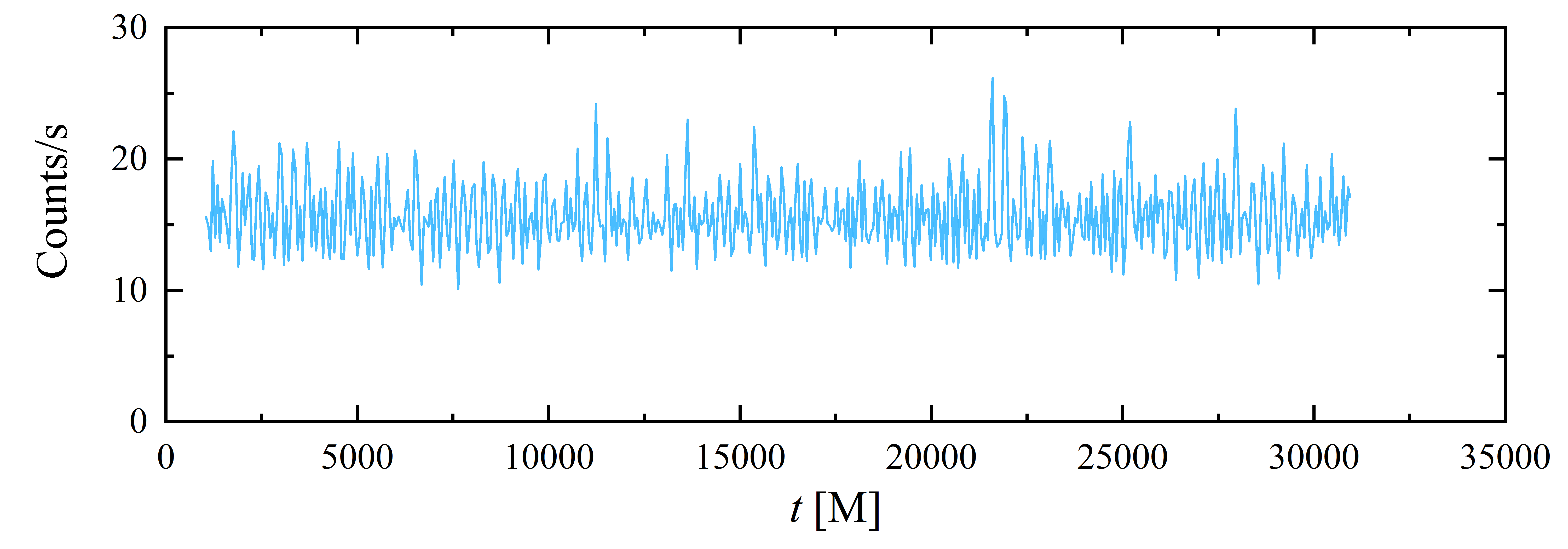}
\includegraphics[width=8.5cm]{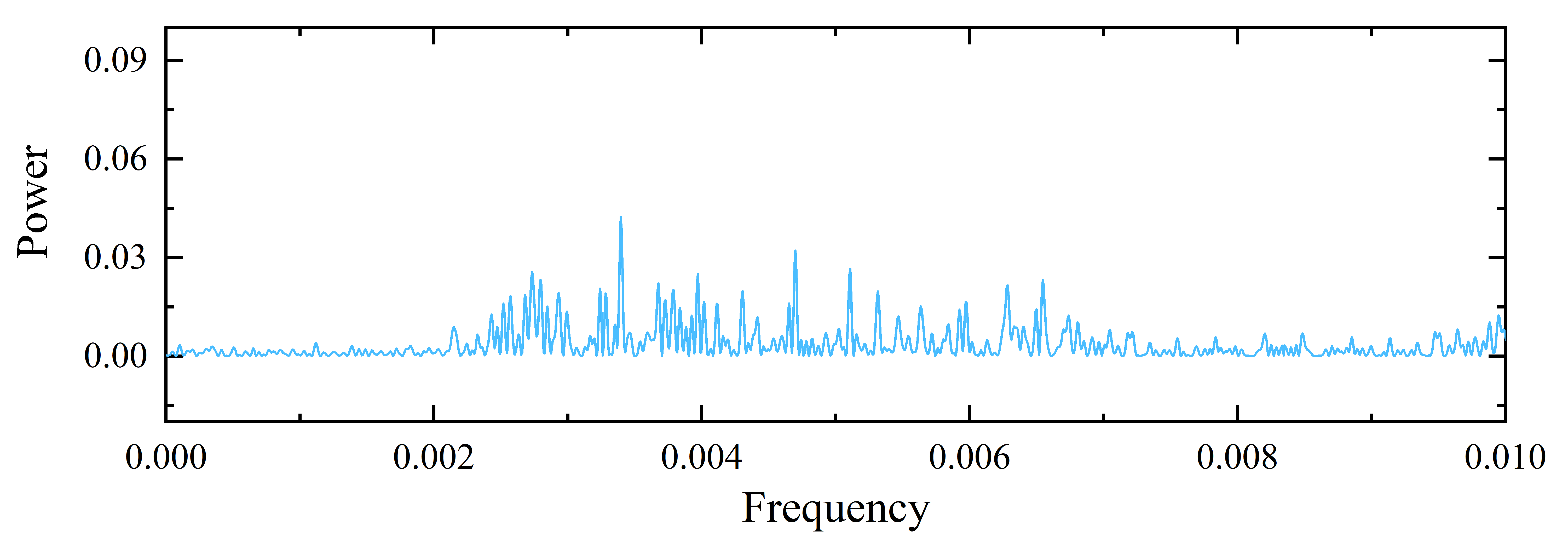}
\caption{(colour online) Light curves (left column) and corresponding power spectra (right column) for orbits 1--6 (top to bottom). For the regular orbits 1 and 2, clear periodicity is evident directly in the light curves. Although no obvious regularity can be discerned from the light curve of orbit 3, its power spectrum exhibits isolated, sharp, narrow peaks---similar to those of orbits 1 and 2---indicating an underlying regular nature. In contrast, the light curves of chaotic orbits 4--6 display large, irregular fluctuations, and their power spectra show continuous, low-amplitude broad peaks resembling mountainous ridges. These results confirm that the regular or chaotic nature of hot-spot orbits can be imprinted in their light curves.}}\label{fig3}
\end{figure*}

\emph{Discussion.}---Currently, effectively detecting chaotic orbits in astrophysics remains an unresolved challenge. In this letter, we demonstrate that, in Kerr spacetime with an external asymptotically uniform electromagnetic field, the light curves of chaotic and regular hot-spots differ significantly. The light curve of an ordered hot-spot exhibits periodic rhythms, akin to a well-organized pattern, while that of a chaotic hot-spot resembles a horizon overgrown with weeds, lacking discernible order. In some cases, regular hot-spots may not show obvious periodicity or quasi-periodicity in their light curves, but their regular nature is still reflected in the power spectrum. Specifically, the power spectra of chaotic hot-spots often exhibit continuous, low-amplitude broad peaks resembling noise, while those of regular hot-spots feature isolated, sharp, narrow peaks. These findings suggest that light curves can serve as a medium for encoding information about chaotic orbits in curved spacetime. Furthermore, given the ubiquity of chaotic orbits in curved spacetime, this effect must be considered in the data processing and analysis of light curves.

It is important to note that in the vacuum spacetime described by general relativity, hot-spots exhibit no chaotic motion, as their equations of motion are integrable. However, three factors can lead to violations of Liouville's integrability theorem in the geodesic equations of hot-spots due to insufficient integrals of motion, thereby inducing chaotic behavior: (i) the spin of the hot-spot, (ii) additional sources in spacetime (e.g., electromagnetic fields or dark matter), and (iii) ``hair'' parameters introduced by modified gravity theories. In other words, if chaotic features in light curves are observed, at least one of the above factors must be present.

We also note that although chaotic signal components have been identified in the light curves of GRS 1915+105 \cite{Misra et al. (2004)} and Hercules X-1 \cite{Voges et al. (1987)} through correlation dimension analysis, this does not imply the detection of chaotic orbits. The origin of these signals remains unknown, and they have not been linked to chaotic dynamics. In contrast, our work is based on simulating light curves derived from chaotic hot-spot orbits, revealing the distinctive features of chaotic trajectories. This not only represents the first attempt to use light curves to observe chaotic orbits, but also provides a potential dynamical framework for understanding the chaotic signals observed in the light curves.

This research has been supported by the National Natural Science Foundation of China [Grant No. 12403081].

\end{document}